\documentclass[reprint,superscriptaddress,amsmath,amssymb,aps,prl]{revtex4-2}

\usepackage{graphicx}
\usepackage{bm}
\usepackage{xcolor}
\usepackage{tikz}
\usepackage[colorlinks=true,linkcolor=blue,urlcolor=blue,citecolor=blue]{hyperref}

\newcommand{\cD}{\mathcal{D}}
\newcommand{\cN}{\mathcal{N}}
\newcommand{\cO}{\mathcal{O}}
\newcommand{\cP}{\mathcal{P}}
\newcommand{\w}{\omega}
\newcommand{\Bw}{\boldsymbol{\w}}
\newcommand{\Bx}{\boldsymbol{x}}
\newcommand{\Bz}{\boldsymbol{z}}
\newcommand{\By}{\boldsymbol{y}}
\newcommand{\Bxi}{\boldsymbol{\xi}}
\newcommand{\RR}{\mathbb{R}}
\newcommand{\Id}{1\!\!1}
\newcommand{\Jt}{\tilde J}
\newcommand{\nn}{\nonumber}
\newcommand{\beq}{\begin{equation}}
\newcommand{\eeq}{\end{equation}}
\newcommand{\of}[1]{\left(#1\right)}
\newcommand{\off}[1]{\left[#1\right]}
\newcommand{\vev}[1]{\left\langle #1\right\rangle}
\newcommand{\vevdots}[1]{\langle #1\,\cdots\rangle}
\newcommand{\Res}{\mathop{\mathrm{Res}}}

\definecolor{pencilblue}{HTML}{2A78D6}
\definecolor{pencilorange}{HTML}{EB6834}
\definecolor{pencilgray}{HTML}{B5B4AE}
\definecolor{pencilink}{HTML}{52514E}
\begin{document}

\title{
Solving spectral flow in AdS$_3$: from covering maps to pencils
}

\author{Sergio Iguri}
\email{siguri@iafe.uba.ar}
\affiliation{Instituto de Astronom\'ia y F\'isica del Espacio (IAFE), CONICET - Universidad de Buenos Aires, Facultad de Ciencias Exactas y Naturales, Ciudad Universitaria, 1428 Buenos Aires, Argentina}

\author{Nicolas Kovensky}
\email{nicolas.kovensky@iflp.unlp.edu.ar}
\affiliation{Instituto de F\'isica de La Plata - CONICET, Diagonal 113 e/ 63 y 64, 1900 La Plata, Argentina}
\affiliation{Departamento de F\'isica, Facultad de Ciencias Exactas y Naturales, Universidad de Buenos Aires, Ciudad Universitaria, 1428 Buenos Aires, Argentina}

\author{Juli\'an H.~Toro}
\email{jtoro@dm.uba.ar}
\affiliation{Instituto de Astronom\'ia y F\'isica del Espacio (IAFE), CONICET - Universidad de Buenos Aires, Facultad de Ciencias Exactas y Naturales, Ciudad Universitaria, 1428 Buenos Aires, Argentina}
\affiliation{Instituto de Investigaciones Matem\'aticas Luis A.~Santal\'o (IMAS), CONICET - Universidad de Buenos Aires, Facultad de Ciencias Exactas y Naturales, Ciudad Universitaria, 1428 Buenos Aires, Argentina}

\begin{abstract}
Strings in AdS$_3$ with pure NS-NS flux are described by the exactly solvable SL(2,$\RR$) WZW model, yet correlators of winding strings, given by spectrally flowed operators, were so far known only up to three points. We obtain all of them, for any number of insertions and windings, in terms of unflowed correlators, proving the four-point conjecture of Dei and Eberhardt. The key ingredient is a pencil of rational maps from the worldsheet to the boundary which exists for generic insertion points, unwinds all insertions at once, and collapses to the covering map when the latter exists.
\end{abstract}

\maketitle

\noindent {\bf \textit{Introduction}}.---Most non-topological examples of holographic duality involve, on the gravity side, string theory on curved backgrounds for which we lack a complete quantization. AdS$_3$/CFT$_2$ with NS-NS flux is a notable exception. It underlies the microscopic counting of black hole (BH) entropy~\cite{Strominger:1996sh}, most directly for the BTZ BH~\cite{Banados:1992wn,Strominger:1997eq}, and the quantum description of the two-dimensional Euclidean BH~\cite{Witten:1991yr}. It also describes a series of explicit microstate geometries~\cite{Martinec:2018nco,Bufalini:2021ndn,Bufalini:2022wyp}. Single-trace $T\bar{T}$ irrelevant deformations of the boundary CFT further lead to a solvable instance of non-AdS holography~\cite{Giveon:2017nie,Chakraborty:2019mdf,Dei:2024sct}.

The worldsheet theory for strings in AdS$_3$ is based on the SL(2,$\RR$) Wess-Zumino-Witten (WZW) model at level $k$, and is under exact control at finite curvature \cite{Kutasov:1999xu,Maldacena:2000hw}. Its spectrum contains the spectrally flowed images of the discrete and continuous representations, i.e.~short and long strings winding around the boundary \cite{Maldacena:2000hw}. The AdS$_3$/CFT$_2$ duality has been established at the tensionless point \cite{Gaberdiel:2018rqv,Eberhardt:2018ouy,Eberhardt:2019ywk}, where correlators vanish unless an appropriate covering map from the worldsheet to the boundary of AdS$_3$ exists. For $k>3$ the dual has been identified, in conformal perturbation theory, with a deformed symmetric orbifold~\cite{Eberhardt:2021vsx,Balthazar:2021xeh}. Twisted sectors of the dual CFT correspond to flowed sectors. Testing and eventually proving the duality at generic $k$ therefore requires computing correlators of spectrally flowed operators.

Such observables are notoriously hard to compute \cite{Maldacena:2001km}. Flowed operators are  not affine primaries, hence Ward identities do not  easily fix their correlators. Three-point functions have been obtained in \cite{Dei:2021xgh,Bufalini:2022toj}. So far, a related four-point formula was conjectured in \cite{Dei:2021yom}, and a specific family of residues in spin space has been computed for all $n$-point functions \cite{Hikida:2023jyc,Knighton:2023mhq,Knighton:2024qxd,Iguri:2026lxp}. Generic correlators have remained out of reach
(see \cite{McStay:2026hlr} for a recent conjecture).

In this Letter we obtain all genus-zero spectrally flowed correlators, which are supported away from the covering-map locus, in terms of the unflowed ones. We do so in the $y$-basis, related to the more familiar $x$-basis by an integral transform \cite{Dei:2021xgh}. In particular, we prove the conjecture of \cite{Dei:2021yom}. Our main tool is a pencil of rational maps, obtained from the linear system that defines covering maps when they exist. Unlike the latter, the pencil exists for generic insertion points. With respect to the resulting deflowed current, all flowed operators become unflowed primaries. For an $n$-point function $F$ with even $\Omega\equiv\sum_i\w_i$, the Ward identities give
\beq
F = \left|\cP_{\Bw}(\Bx,\Bz)^{-k}\, \prod_{i=1}^n \of{\frac{d\xi_i}{dy_i}}^{j_i}\right|^2 \Phi(\Bxi,\Bz) \, ,
\label{L: general solution}
\eeq
where $\Bx$ and $\Bz$ denote the boundary and worldsheet insertion points, $\xi_i$ are the members of the pencil selected by the values of $y_i$, $\Phi$ satisfies the global Ward identities of unflowed primaries of the same spins, and $\cP_{\Bw}$ is proportional to the determinant of that linear system.  In terms of $\Phi$, the Knizhnik-Zamolodchikov (KZ) equations reduce to the unflowed ones, and fixing the normalization identifies $\Phi$ with the unflowed correlator. For odd $\Omega$, the linear system includes one additional condition. The conclusion is analogous, although one of the unflowed operators appears with spin $\tilde{\jmath} = \frac k2-j$. A similar statement holds for the edge cases where $\Omega=2\w_{\rm max}-d$ with $\w_{\rm max}=\max_i\w_i$ and $2\leq d\leq n-2$, although an additional coincidence limit must be imposed on the unflowed correlator \cite{Toro:general}.

\vspace{1.5mm}\noindent {\bf \textit{The pencil}}.---We consider $y$-basis vertex operators $V_i \equiv V^{\w_i}_{j_i}(x_i,y_i,z_i)$ of spins $j_i$ and spectral flow charges $\w_i \geq 0$, at boundary points $x_i$ and worldsheet points $z_i$, in the conventions of~\cite{Dei:2021xgh,Dei:2021yom}, omitting the antiholomorphic dependence. The $V_i$ are characterized by
\begin{subequations}
\label{L: OPEs}
\begin{equation}
J^+(z)\,V_i =
\frac{\partial_{y_i} V_i}{(z-z_i)^{\w_i+1}}
+ \sum_{m=0}^{\w_i-1} \frac{J^+_m V_i}{(z-z_i)^{m+1}}
+ \cdots \, ,
\end{equation}
\begin{equation}
J^3(x_i,z)\,V_i =
\frac{\big(y_i\,\partial_{y_i} + j_i + \frac{k\w_i}{2}\big)V_i}{z-z_i}
+ \cdots \, ,
\end{equation}
\begin{equation}
J^-(x_i,z)\,V_i =
(z-z_i)^{\w_i-1}\,\big(y_i^2\partial_{y_i} + 2j_iy_i\big)V_i
+ \cdots \, ,
\end{equation}
\end{subequations}
where $J(x;z) \equiv J^-(x,z) = J^- - 2xJ^3 + x^2J^+$ and $J^3(x,z) = J^3 - x\,J^+$. The modes $J^+_mV_i$, $1\leq m\leq\w_i-1$, are not fixed. For $\w = 0$, $V^0_j(x,y,z) = V_j(x+y,z)$.

To deal with all insertions at once, consider maps $\Gamma = p/q$ from the worldsheet to the boundary, with $p$ and $q$ polynomials of degree at most $D$ such that $\Gamma(z) - x_i = \cO[(z-z_i)^{\w_i}]$, i.e.
\beq
[p-x_iq]_m(z_i) = 0 \, , \qquad m = 0,\dots,\w_i-1 \, ,
\label{L: linear system}
\eeq
where $[\cdots]_m(z_i)$ is the Taylor coefficient of order $m$ at $z_i$. Solutions without further critical points are covering maps~\cite{Lunin:2000yv}, which for $n\geq4$ exist only on a locus of codimension $n-3$. There are $2D+2$ unknowns and $\Omega$ conditions. For even $\Omega$ the system is square at $D=N\equiv\frac12\Omega-1$, with determinant $\tau_{\Bw}(\Bx,\Bz)$, which does not vanish identically if and only if $\Omega\geq2\w_{\rm max}$. Below this bound the correlator either vanishes~\cite{Maldacena:2001km,Eberhardt:2019ywk} or belongs to the edge cases discussed below. When the bound holds, for generic insertions there is no non-zero solution at degree $N$. One degree higher, the solutions form a plane spanned by $(p_1,q_1)$ and $(p_2,q_2)$, i.e.~a \textit{pencil} of rational maps of degree $N+1$,
\beq
\Gamma_s(z) = \frac{p_s(z)}{q_s(z)} \, , \quad p_s = s\,p_1 + p_2 \, , \quad q_s = s\,q_1 + q_2 \, .
\label{L: pencil}
\eeq
The coordinate $s$ depends on the choice of basis, under which all statements below are covariant. When $\tau_{\Bw}=0$ the pencil collapses to a single map. For $n=4$ this happens on the  covering-map locus. For instance, $\tau_{(1,1,1,1)} = x_{13}x_{24}z_{12}z_{34} - x_{12}x_{34}z_{13}z_{24}$ vanishes precisely when the boundary and worldsheet cross-ratios coincide.
For $n\geq5$, the hypersurface $\tau_{\Bw}=0$ contains the covering-map locus, if any.

Since $p_1q_2-p_2q_1$ has degree at most $\Omega$ and vanishes to order $\w_i$ at each $z_i$,
\beq
p_1(z)\,q_2(z) - p_2(z)\,q_1(z) = C\prod_{i=1}^n (z-z_i)^{\w_i} \equiv R(z) \, ,
\label{L: area form}
\eeq
with $C\neq0$ for $\tau_{\Bw}\neq0$. We also define
\beq
\cP_{\Bw}(\Bx,\Bz) = \prod_{i<k} z_{ik}^{-\frac12\w_i\w_k}\;\tau_{\Bw}(\Bx,\Bz) \, , \qquad z_{ik} = z_i - z_k \, ,
\label{L: P}
\eeq
which transforms as a product of primaries of weights $-\frac12\w_i$ at $x_i$ and $\frac14\w_i^2$ at $z_i$.

\begin{figure}[t]
\centering
\begin{tikzpicture}[scale=0.62]
\fill[pencilblue!7] (0,0) -- (10.6,0) -- (12.6,2.8) -- (2,2.8) -- cycle;
\draw[thick,pencilblue] (0,0) -- (10.6,0) -- (12.6,2.8) -- (2,2.8) -- cycle;
\node[pencilblue] at (11.2,0.45) {$s$};
\foreach \x/\y/\i/\h/\lab in {2.1/0.9/1/3.25/{a_1(s)=y_1},4.6/2.0/2/4.05/{a_2(s)=y_2},6.9/0.8/3/3.25/{a_3(s)=y_3},10.2/1.7/n/4.05/{a_n(s)=y_n}}{
  \fill[pencilorange] (\x,\y) circle (3.6pt);
  \node[below] at (\x,\y-0.1) {$\xi_{\i}$};
  \draw[pencilgray] (\x,\y+0.12) -- (\x,\h);
  \node[above,font=\footnotesize] at (\x,\h) {$\lab$};
}
\node[pencilink] at (8.6,1.25) {$\cdots$};
\end{tikzpicture}
\caption{The value of $y_i$ at $z_i$ selects a member $\xi_i$ of the pencil.}
\label{fig: pencil}
\end{figure}
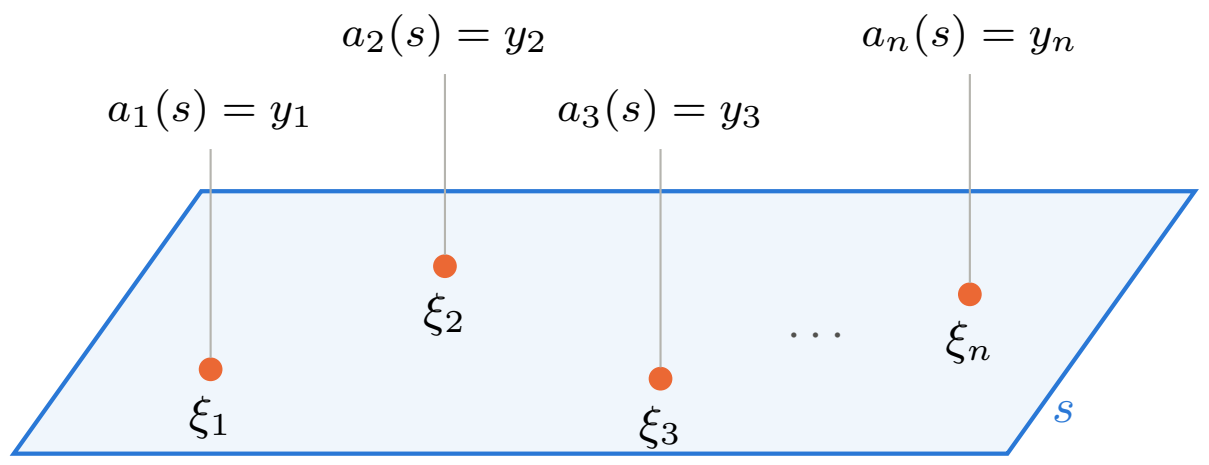

Near each $z_i$, $\Gamma_s(z) = x_i + (z-z_i)^{\w_i}A_i(s;z)$, where $A_i$ is a M\"obius transformation of $s$, regular and invertible at $z_i$. Its value $a_i(s) = A_i(s;z_i)= [p_s-x_iq_s]_{\w_i}(z_i)/q_s(z_i)$ is the local coefficient of the member $s$. The variable $y_i$ selects the member $\xi_i$ with $a_i(\xi_i)=y_i$, see Fig.~\ref{fig: pencil}, which for an unflowed operator is the one with $\Gamma_{\xi_i}(z_i)=x_i+y_i$. By Eq.~\eqref{L: area form},
\beq
\frac{d\xi_i}{dy_i} = \frac{1}{a_i'(\xi_i)} =\frac{q_{\xi_i}(z_i)^2}{C\,\pi_i} \, , \qquad \pi_i \equiv \prod_{k\neq i} z_{ik}^{\,\w_k} \, ,
\label{L: jacobian}
\eeq
and, since $a_i$ is M\"obius, any two members satisfy
\beq
\big(a_i(s_1) - a_i(s_2)\big)^2 = a_i'(s_1)\,a_i'(s_2)\,(s_1-s_2)^2 \, .
\label{L: Mobius identity}
\eeq
The dependence on the insertion points is controlled by
\beq
W = \frac{q_1q_2' - q_1'q_2}{R} \, , \qquad K(s;z) = \frac{\partial_z\Gamma_s}{\partial_s\Gamma_s}  \, ,
\label{L: W and K}
\eeq
where $K(s;z)= K^- - 2sK^3 + s^2K^+$. In the End Matter  we show  that
\begin{subequations}
\label{L: motion}
\begin{align}
\partial_{x_i}\xi_\ell &= \Res_{z=z_i}\frac{q_{\xi_\ell}(z)^2}{R(z)\,(z-z_\ell)} + \varphi_i(\xi_\ell) \, , \label{L: motion x}\\
\partial_{z_i}\xi_\ell &= -\Res_{z=z_i}\frac{K(\xi_\ell;z)}{z-z_\ell} - \psi_i(\xi_\ell) \, , \label{L: motion z}
\end{align}
\end{subequations}
at fixed $y_\ell$, with $\varphi_i$ and $\psi_i$ quadratic polynomials independent of $\ell$, and that
\begin{subequations}
\label{L: tau}
\begin{align}
\Res_{z=z_i} W(z) &= \partial_{x_i}\log\cP_{\Bw} \, , \label{L: tau x}\\
\Res_{z=z_i}\det K(z) &= -\partial_{z_i}\log\cP_{\Bw} \, , \label{L: tau z}
\end{align}
\end{subequations}
with $\det K \equiv K^+K^- - (K^3)^2$.

\vspace{1.5mm}\noindent {\bf \textit{Deflowed current}}.---We define
\beq
\Jt(s;z) = \frac{J\big(\Gamma_s(z);z\big) + k\,\partial_z\Gamma_s(z)}{\partial_s\Gamma_s(z)} = \Jt^- - 2s\,\Jt^3 + s^2\Jt^+ \, .
\label{L: deflowed current}
\eeq
For a single operator we can take $\Gamma_s = x_i + s(z-z_i)^{\w_i}$, and, in terms of modes at $z_i$, this gives $\Jt^\pm_m = J^\pm_{m\pm\w_i}(x_i)$ and $\Jt^3_m = J^3_m(x_i) - \frac{k\w_i}{2}\delta_{m,0}$, which is the spectral flow automorphism~\cite{Maldacena:2000hw}. By Eqs.~\eqref{L: OPEs}, $V_i$ is annihilated by the positive modes of $\Jt$, while
\beq
\Jt^+_0 = \partial_{y_i} \, , \quad \Jt^3_0 = y_i\,\partial_{y_i} + j_i \, , \quad \Jt^-_0 = y_i^2\partial_{y_i} + 2j_iy_i \, .
\label{eq: modes single Jtilde}
\eeq
With respect to $\tilde{J}$, $V_i$ is an unflowed primary of spin $j_i$ at the boundary point $y_i$. For the pencil, inserting $\Gamma_s = x_i + (z-z_i)^{\w_i}A_i$ in Eq.~\eqref{L: deflowed current} gives the exact identity
\beq
\Jt(s;z) = \frac{\Jt_i\big(A_i(s;z);z\big)}{\partial_sA_i(s;z)} + k\,\frac{\partial_zA_i(s;z)}{\partial_sA_i(s;z)} \, ,
\label{L: composition}
\eeq
where $\Jt_i(\sigma;z)$ is the deflowed current of the single-operator map $x_i + \sigma(z-z_i)^{\w_i}$. Near each insertion point, $\Jt(s;z)$ coincides with $\Jt_i\big(a_i(s);z\big)/a_i'(s)$ up to terms that are regular at $z_i$, so that
\beq
\Res_{z=z_i}\vevdots{\Jt(s;z)} = \frac{\cD_{y_i}\big(a_i(s)\big)\,F}{a_i'(s)} \, ,
\label{L: residues}
\eeq
with $\cD_{y}(\sigma) = (\sigma-y)^2\partial_{y} - 2j(\sigma-y)$. All vertex operators thus behave simultaneously as unflowed primaries with respect to $\Jt$. The latter satisfies the same algebra as $J$, and its Sugawara tensor $\tilde T$, defined by $(k-2)\,T = \frac12(J^+J^-) + \frac12(J^-J^+) - (J^3J^3)$ with $J \to \Jt$, also sees them as unflowed primaries. By Eq.~\eqref{L: area form}, $T - \tilde T$ is linear in the currents, and comparing double poles gives
\beq
T = \tilde T - \big(K^+\Jt^- + K^-\Jt^+ - 2K^3\Jt^3\big) + k\,\det K \, .
\label{L: T vs Tt}
\eeq
For a single operator, $K = \w_i s/(z-z_i)$ and Eq.~\eqref{L: T vs Tt} is the spectral flow of the Virasoro generators, $L_m = \tilde L_m - \w_i\,\Jt^3_m - \frac k4\,\w_i^2\,\delta_{m,0}$~\cite{Maldacena:2000hw}.

\vspace{1.5mm}\noindent {\bf \textit{Ward identities and KZ equations}}.---By Eq.~\eqref{L: deflowed current}, $\vevdots{\Jt(s;z)}$ decays as $z^{-2}$, hence the sum of the residues \eqref{L: residues} must vanish. Writing $F$ as in Eq.~\eqref{L: general solution} (with $\Phi$ allowed to depend on $\Bx$ for now) Eq.~\eqref{L: Mobius identity} turns the residues into $\cD_{\xi_i}(s)$ acting on $\Phi$, giving
\beq
\sum_{i=1}^n \cD_{\xi_i}(s)\,\Phi = 0 \qquad \text{for all } s \, ,
\label{L: global Ward}
\eeq
i.e.~the global Ward identities of unflowed primaries of spins $j_i$ at  $\xi_i$. The dependence on $\Bx$ follows from $J^+_0V_i = \partial_{x_i}V_i$. Inverting Eq.~\eqref{L: deflowed current},
\beq
J^+ = \chi^+ \Jt^- + \chi^- \Jt^+ - 2\chi^3 \Jt^3-kW \, ,
\label{L: Jplus}
\eeq
where $\chi(s;z) = q_s^2/R= \chi^- - 2s\chi^3 + s^2\chi^+$, so that
\beq
G^+(z) = \off{\sum_{\ell=1}^n \frac{\Phi^{-1}\big(\chi_\ell\,\partial_{\xi_\ell} + j_\ell\,\partial_{\xi_\ell}\chi_\ell\big)\Phi}{z-z_\ell} - k W}F  \, ,
\label{L: Gplus}
\eeq
with $G^+=\vev{J^+(z)\cdots}$ and $\chi_\ell = \chi(\xi_\ell;z)$. By Eqs.~\eqref{L: motion x} and \eqref{L: tau x}, the conditions $\Res_{z=z_i}G^+ = \partial_{x_i}F$ hold if and only if,  at fixed $\Bxi$, $\Phi$ does \textit{not} depend on $\Bx$. The residue in Eq.~\eqref{L: motion x} accounts for the motion of the points, its $\xi_\ell$-derivative for that of the Jacobians, the terms with $\varphi_i$ cancel by Eq.~\eqref{L: global Ward}, and Eq.~\eqref{L: tau x} accounts for $\cP_{\Bw}^{-k}$. The KZ equations~\cite{Knizhnik:1984nr} give $\partial_{z_i}F = \Res_{z=z_i}\vevdots{T(z)}$ at fixed $\Bx$ and $\By$ and work in the same way: by Eq.~\eqref{L: T vs Tt}, $\tilde T$ gives the unflowed KZ operator acting on $\Phi$, the term linear in $\Jt$ the motion~\eqref{L: motion z}, and $k\det K$ the derivative of $\cP_{\Bw}^{-k}$, by Eq.~\eqref{L: tau z}. Hence the flowed KZ equations hold iff
\beq
(k-2)\,\partial_{z_i}\Phi = \sum_{\ell\neq i}\frac{\cD^+_i\cD^-_\ell + \cD^-_i\cD^+_\ell - 2\,\cD^3_i\cD^3_\ell}{z_i-z_\ell}\,\Phi \, ,
\label{L: KZ}
\eeq
with $\cD_{\xi_i}(s) = \cD_i^--2s\cD_i^3+s^2\cD_i^+$, i.e.~$\Phi$ satisfies the unflowed KZ equations at the points $\xi_i$.

\vspace{1.5mm}\noindent {\bf \textit{Normalization}}.---In summary, for even $\Omega \geq 2\w_{\rm max}$ and $\tau_{\Bw}\neq0$, $F$ satisfies all Ward identities and KZ equations if and only if it takes the form~\eqref{L: general solution}, with $\Phi$ a solution of the unflowed global Ward identities and KZ equations at the points $\xi_i$.
Let $X_{ik} = \sqrt{z_{ik}}\,\cP_{ik}$, with $\cP_{ik}$ given by Eq.~\eqref{L: P} for the charges $\Bw+e_i+e_k$ and the conditions $a_i = y_i$, $a_k = y_k$ added to the system. Then, up to a phase
\beq
X_{ik} = \cP_{\Bw}\,\frac{\xi_i - \xi_k}{\sqrt{\xi_i'\,\xi_k'}} \, , \qquad \xi_i' \equiv \frac{d\xi_i}{dy_i} \, ,
\label{L: X}
\eeq
and Eq.~\eqref{L: general solution} becomes
\beq
F = |\cP_{\Bw}^{J-k}\prod_{i<k}X_{ik}^{\,n_{ik}}|^2F_0\big(\text{{\small cross-ratios of }}X_{ik},\Bz\big) \, ,
\label{L: F in X}
\eeq
with $J=\sum_ij_i$ and $\sum_{k\neq i}n_{ik} = -2j_i$. For $n = 3,4$ this is the structure found in \cite{Dei:2021xgh,Bufalini:2022toj,Dei:2021yom}, which suggests
\beq
\Phi(\Bxi,\Bz) = \vev{\prod_{i=1}^n V_{j_i}(\xi_i,z_i)} \, .
\label{L: normalization}
\eeq
This holds trivially  in the unflowed case, where $\Gamma_s(z)=s$ and $\xi_i=x_i+y_i$. We prove it in general using three known facts. (i) The series identification~\cite{Maldacena:2000hw,Bufalini:2022toj},
\beq
V^\w_j(x,0,z) = \cN(j)\,\lim_{y\to\infty} y^{k-2j}\,V^{\w+1}_{\frac k2-j}(x,y,z) \, ,
\label{series id}
\eeq
with $\cN(j)^2 = B(j)/B\of{\frac k2-j}$, where $B(j)$ normalizes the unflowed two-point function. (ii) The point-splitting formula \cite{Iguri:2022eat}
\beq
V^\w_j(x,y,z) = \lim_{\epsilon\to0}\epsilon^{j\w}\,V_j(x+y\,\epsilon^\w,z+\epsilon)\,V^\w_0(x,z)\,.
\label{L: fusion}
\eeq
(iii) The unflowed OPE of $V_j$ with $V_{\frac{k}{2}}$ contains only $V_{\frac k2-j}$ \cite{Fateev,Maldacena:2001km}.

By (i), $V^1_0\sim V_{\frac k2}$, and by (i) and (ii) every flowed correlator can be obtained as a limit of an unflowed one with $\Omega$ additional insertions of $V^1_0$. Before the limit, this is a flowed correlator with unit charges at the $V^1_0$, of the form~\eqref{L: general solution}. Since the $V^1_0$ have spin zero and no $y$-dependence, $\Phi$ depends neither on their points $\xi_a$ nor, by Eq.~\eqref{L: KZ}, on their $z_a$, in the region $\tau_{\Bw}\neq0$, which is connected and contains coincident boundary points at distinct worldsheet points. By (iii), pairs of $V^1_0$ at different boundary points annihilate as $z_{ab}\to0$, i.e.~$V^1_0(x_a,z_a)\,V^1_0(x_b,z_b)= x_{ab}^{-k}z_{ab}^{\frac k2}\off{\Id + \cO(z_{ab})}$, while the pencil reduces to that of the remaining charges and $\cP_{\Bw}$ acquires the factor $x_{ab}\,z_{ab}^{-\frac12}$. Once all charges vanish, Eq.~\eqref{L: normalization} holds. The limits that build the flowed operators leave $\Phi$ unchanged. As $V^1_0(x,z+\epsilon)$ merges with $V^{\w-1}_0(x,z)$, the pencil acquires ramification $\w$ at $z$ and $\cP_{\Bw}^{-k}\sim\epsilon^{-\frac k2(\w-1)}$. As $V_j(x+y\,\epsilon^\w,z+\epsilon)$ approaches $V^\w_0(x,z)$, the pencil is unchanged, $V_j$ goes to the member with local coefficient $y$, and its Jacobian scales as $\epsilon^{-j\w}$. Eq.~\eqref{L: fusion}, with $j=\frac k2$ and $y=0$ in the first case, compensates both factors, which proves Eq.~\eqref{L: normalization} for flowed operators normalized as in~\cite{Dei:2021xgh,Dei:2021yom}.

\vspace{1.5mm}\noindent {\bf \textit{Odd parity}}.---For odd $\Omega\geq2\w_{\rm max}-1$ the system~\eqref{L: linear system} is never square, and at degree $\frac12(\Omega+1)$ the solution space is, in general, three-dimensional. We remove one degree of freedom by imposing, for a chosen operator $V_r$
\beq
[p-x_rq]_{\w_r}(z_r) = y_r\,q(z_r) \, .
\label{L: local yr}
\eeq
The choice of $r$ is immaterial as long as $\Omega\geq2\w_r+1$.
The system~\eqref{L: linear system}--\eqref{L: local yr} is square at degree $\frac12(\Omega-1)$, with determinant $\tau_r$. When $\tau_r\neq 0$, its solutions at one degree higher form a pencil, for which we keep the notation of Eqs.~\eqref{L: pencil}--\eqref{L: Mobius identity}. At $y_r=0$ this is the even pencil with $\w_r\to\w_r+1$, but the construction works for any value of $y_r$.
Eq.~\eqref{L: area form} reads $R=C(z-z_r)\prod_i(z-z_i)^{\w_i}$, so that the $\pi_i$ are computed with the charges $\Bw+e_r$, and we define $\cP_r$  by Eq.~\eqref{L: P} for these charges, with $\tau_r$ instead of $\tau_{\Bw}$. The operators $V_i$ with $i\neq r$ behave as before. At $z_r$ all members share the local coefficient $y_r$, and differ only at the next order,
\beq
\Gamma_s(z) = x_r + (z-z_r)^{\w_r}\big[y_r + (z-z_r)\,\tilde A_r(s;z)\big] \, ,
\label{L: Gamma near zr}
\eeq
with $\tilde A_r$ M\"obius in $s$, and, Eq.~\eqref{L: area form}, $\tilde a_r(s) = \tilde A_r(s;z_r)$ satisfies $\tilde a_r'(s) = C\pi_r/q_s(z_r)^2$.

Eq.~\eqref{L: composition} holds near $z_r$ with $\tilde A_r$ in place of $A_i$, and with $\Jt_r(\sigma;z)$ the deflowed current given by the map $x_r + (z-z_r)^{\w_r}\big[y_r+\sigma(z-z_r)\big]$, hence
\begin{subequations}
\label{L: Jt r}
\begin{equation}
\Jt^+_r = (z-z_r)^{\w_r+1}J^+ \, ,
\end{equation}
\begin{equation}
\Jt^3_r = J^3(x_r,z) - y_r(z-z_r)^{\w_r}J^+ - \frac{k(\w_r+1)}{2(z-z_r)} \, ,
\end{equation}
\begin{equation}
\Jt^-_r = \frac{J^-\big(x_r+y_r(z-z_r)^{\w_r},z\big)}{(z-z_r)^{\w_r+1}} + \frac{k\w_r y_r}{(z-z_r)^2} \, .
\end{equation}
\end{subequations}
By Eqs.~\eqref{L: OPEs}, $\Jt^+_r$ is regular at $z_r$, $\Jt^3_r$ has a simple pole with residue $(j_r-\frac k2)V_r$, and the double poles of $\Jt^-_r$ cancel, leaving a simple pole whose residue is not fixed.
It follows that $V_r$ is seen as an unflowed primary of spin $\tilde \jmath_r = \frac k2-j_r$ at $\tilde y=-1/\sigma=0$, the zero mode of $\Jt^-_r$ playing the role of $\partial_{\tilde y}$. For $y_r=0$, $\Jt_r$ is the spectral flow by $\w_r+1$ units, which gives the series identification~\eqref{series id} at the level of the current. In the pencil, $V_r$ therefore sits at the member $\xi_r$ where $\tilde a_r$ has its pole, i.e.~$q_{\xi_r}(z_r)=0$, the only member for which Eq.~\eqref{L: local yr} becomes trivial. With $\tilde y_r(s) = -1/\tilde a_r(s)$, Eq.~\eqref{L: residues} holds at $z_r$ with $a_r(s)$, $y_r$ and $j_r$ replaced by $\tilde y_r(s)$, $\tilde y_r(\xi_r)=0$ and $\tilde \jmath_r$, and $d\xi_r/d\tilde y_r = C\pi_r/q_1(z_r)^2$, where $q_s(z_r) = q_1(z_r)(s-\xi_r)$.

The rest of the analysis applies to this pencil. Eqs.~\eqref{L: motion}, \eqref{L: tau} and \eqref{L: T vs Tt} hold for it, with $\cP_r$ instead of $\cP_{\Bw}$, the derivatives of $\xi_r$ being taken at fixed $\tilde y_r$. Eqs.~\eqref{L: motion x} and \eqref{L: tau x} also hold with $x_r$ replaced by $y_r$, and $\varphi_r$ by another quadratic polynomial, if the residues at $z_r$ are multiplied by $(z-z_r)^{\w_r}$, which accounts for the pole of order $\w_r+1$ of $J^+$ at $z_r$, with coefficient $\partial_{y_r}F$. Setting $(\tilde y_i,\tilde \jmath_i) = (y_i,j_i)$ for $i\neq r$, the general solution is again Eq.~\eqref{L: general solution}, with $\Phi$ independent of $\Bx$ and $y_r$ at fixed $\xi_i$,  and satisfying the unflowed Ward identities and KZ equations for the spins $\tilde{\jmath}_i$.

We fix $\Phi$ by setting $y_r=0$ and using the even-parity result. By Eq.~\eqref{series id}, $V_r|_{y_r=0} = \cN(j_r)\lim_{y\to\infty}y^{2\tilde{\jmath}_r}\,V^{\w_r+1}_{\tilde \jmath_r}(x_r,y,z_r)$. In this limit the member selected by $y$ tends to $\xi_r$, and $(y^{2}d\xi/dy)^{\tilde \jmath_r}\to(d\xi_r/d\tilde y_r)^{\tilde \jmath_r}$. Eqs.~\eqref{L: general solution} and \eqref{L: normalization} then give, for all $y_r$,
\beq
F = \cN(j_r)\left|\cP_r^{-k}\prod_{i=1}^n\of{\frac{d\xi_i}{d\tilde y_i}}^{\tilde \jmath_i}\right|^2 \vev{\prod_{i=1}^nV_{\tilde \jmath_i}(\xi_i,z_i)} \, .
\label{L: odd solution}
\eeq
Eq.~\eqref{L: odd solution} can also be obtained by inserting the identity written in terms of $\lim_{y\to\infty}y^k\,V^1_{\frac k2}(y)$.  Eqs.~\eqref{L: general solution} and \eqref{L: normalization} then give an unflowed correlator with an extra $V_{\frac k2}$ whose collision with $V_{j_r}$ produces the above spin flip.

In the odd case, we define $X_{(ir)} \equiv X_i = \sqrt{z_{ir}}\,\cP_i$, where $\cP_i$ is given by Eq.~\eqref{L: P} for the charges $\Bw+e_i$,  with the extra condition $a_i=y_i$ and without \eqref{L: local yr}, and similarly for $X_{(ik)} \equiv X_{ikr} = \sqrt{z_{ik}}\,\cP_{ikr}$ when $r\notin\{i,k\}$.
As in Eq.~\eqref{L: X}, these satisfy $X_{(ik)} = \cP_r\,\xi_{ik}/\sqrt{\xi_i'\xi_k'}$ up to a phase, with $\xi_r' = d\xi_r/d\tilde y_r$, giving
\beq
F = \cN(j_r)|\cP_r^{\tilde{J}-k}\prod_{i<k}X_{(ik)}^{\,n_{ik}}|^2F_0(\text{{\small cross-ratios of }} X_{(ik)},\Bz),
\nn
\eeq
with $\tilde{J}=\sum_i\tilde \jmath_i$ and $\sum_{k\neq i}n_{ik} = -2\tilde \jmath_i$. For $n=4$ and $V_{r=3}$  at infinity,
this is the conjecture of~\cite{Dei:2021yom}.

\vspace{1.5mm}\noindent {\bf \textit{Edge cases}}.---For $\Omega=2\w_{\rm max}-d$ with $2\leq d\leq n-2$, $\tau_{\Bw}$, or $\tau_r$ for odd $\Omega$, vanishes identically, but the correlator does not~\cite{Maldacena:2001km,Dei:2021yom}. Here the operator with largest charge, say $V_i$, winds more than all the others together, and a map with $\Gamma(z)-x_i=\cO[(z-z_i)^{\w_{\rm max}}]$ must either have degree at least $\w_{\rm max}$ or be constant. Below this degree, the solutions of Eq.~\eqref{L: linear system} are pairs $(p,q)=(x_iq,q)$ describing the constant map $\Gamma(z)=x_i$, which cannot help unwind the other operators. In order to do so, one needs maps of degree $\w_{\rm max}$, for which Eq.~\eqref{L: linear system} leaves $d+2$ free parameters. As in the odd case, we remove $d$ of them imposing $d$ additional conditions analogous to \eqref{L: local yr}. The system is then square at degree $\w_{\rm max}-1$, generically with non-vanishing determinant. One degree higher, its solutions form a pencil.

The analysis of the odd case carries over to each of these $d$ operators, which become unflowed primaries of spins $\tilde\jmath_r=\frac k2-j_r$, located at the unique member whose $q_s(z)$ vanishes at their $z_r$, namely the constant map discussed above. For $d\geq2$ the KZ equations no longer fix $\Phi$, since they involve the $d$ unknown zero modes separately. However, $\Phi$ does not depend on the $y_r$, and at $y_r=0$ the series identification can be used for each of the $d$ operators. The result takes the form of Eq.~\eqref{L: odd solution}, with $\cP_r$ computed from this system and a factor $\cN(j_r)$ and a reflected spin for each of the $d$ operators, which now sit at the same point of the pencil. As in the odd case, the same conclusion can be obtained by inserting a series of identity operators, each written as a limit of $V^1_{\frac k2}$.

Since unflowed correlators are singular at coincident boundary points, what survives is the coefficient of a definite power of their separation. This can be evaluated using the $H_3^+$--Liouville correspondence~\cite{Ribault:2005wp}, and the result is an integral of a Liouville correlator with $n-d-2$ degenerate fields~\cite{Ribault:2005ms,Toro:general}. For $n=4$, this reproduces the edge-case results of~\cite{Dei:2021yom,Dei:2022pkr}, see the End Matter.

\vspace{1.5mm}\noindent {\bf \textit{Discussion}}.---We have shown that all  genus-zero correlators of $y$-basis spectrally flowed operators in the SL(2,$\RR$) WZW model can be expressed in terms of unflowed correlators of the same spins, evaluated at the points of a pencil of rational maps, with one spin $j\to\frac k2-j$ for odd $\Omega$, and $d$ flipped spins at a common point in the edge cases. The only remnants of spectral flow are the change of variables $y_i\to\xi_i$ and the factor $\cP_{\Bw}^{-k}$. This proves the conjecture of~\cite{Dei:2021yom} and extends it to any number of insertions, with all constants fixed. It also gives its ingredients a geometric meaning: up to Jacobians, the generalized differences $X_{ij}$ of~\cite{Dei:2021xgh,Dei:2021yom} encode the distances $\xi_i-\xi_j$ between members of the pencil, and $X_\emptyset\sim\cP_{\Bw}$ captures the determinant of the linear system of constraints.  At generic $k$, the pencil takes over the role that covering maps play at the tensionless point.

A formula of the same type, built from Pad\'e approximants,  was recently conjectured in~\cite{McStay:2026hlr} for even $\Omega$. As Eq.~\eqref{L: linear system} is itself a multipoint Pad\'e problem, we expect the two constructions to be closely related.

At genus zero, the worldsheet correlators reduce to the pencil and the $H_3^+$ model, which governs unflowed correlators and is under full analytic control~\cite{Teschner:1997ft,Teschner:1999ug}, its $n$-point functions being expressible through Liouville theory~\cite{Ribault:2005wp}. Our method should help study factorization and derive the spacetime OPE~\cite{Iguri:2024yhb,Sriprachyakul:2025ubx}, including all single-cycle flowed channels as well as the multi-cycle channels left open in~\cite{Maldacena:2001km,Iguri:2026lxp}.

Our results allow for a systematic study of the analytic structure of string correlators in AdS$_3$. When the $j$-constraint holds, correlators localize on the  covering-map locus~\cite{Eberhardt:2019ywk}. In our language, this must be encoded in the collapse of the pencil \cite{Toro:general}. More generally, there is a two-parameter lattice of poles in the space of the unflowed spins.  So far, only the series of singularities tied to covering maps and to insertions of the twist-two marginal operator deforming the boundary CFT has been analyzed~\cite{Hikida:2023jyc,Knighton:2024qxd,Iguri:2026lxp}. A second series of poles, related to the previous one by an analogue of the $b\to1/b$ duality of Liouville theory, was pointed out in~\cite{Eberhardt:2021vsx} but remains unexplored. It can now be addressed for any $n$.

The most important open problem is the dual CFT itself. For $k>3$ it is known only in conformal perturbation theory around the symmetric orbifold~\cite{Eberhardt:2021vsx,Balthazar:2021xeh}, and a non-perturbative definition is lacking~\cite{Seiberg:1999xz,Eberhardt:2019qcl}. As in Liouville theory~\cite{Zamolodchikov:1995aa}, the exact correlators obtained here, together with their residues, might suffice to define it. Finally, spectrally flowed correlators are the missing ingredient in several related settings. In the supersymmetric AdS$_3\times S^3\times T^4$ background the bosonic SL(2,$\RR$) model appears at level $k+2$, hence our results give access to correlators of twisted-sector chiral primaries beyond three points~\cite{Iguri:2022pbp,Barone:2025vww,Knighton:2026wva}. On the cigar SL(2,$\RR$)/U(1)~\cite{Witten:1991yr} winding is not conserved and winding modes descend from flowed operators, so that one obtains all genus-zero correlators of the winding condensate behind the FZZ duality~\cite{Hikida:2008pe} and the black hole/string transition~\cite{Chen:2021dsw}. Similarly, they extend the three-point functions computed in the microstate geometries of BTZ black holes~\cite{Martinec:2018nco,Bufalini:2022wyp} to any $n$. Since the single-trace $T\bar T$ spectrum is built from long strings, our results give access to its $n$-point functions~\cite{Giveon:2017nie,Chakraborty:2019mdf,Dei:2024sct}.

\vspace{1.5mm}\noindent {\bf \textit{Note added}}.---While this work was being completed, Ref.~\cite{McStay:2026hlr} appeared, containing a conjecture that partially overlaps with our formulas. Our results were obtained independently and by different methods.

\vspace{1.5mm}\noindent {\bf \textit{Acknowledgments}}.---The work of the authors is supported by CONICET-Argentina and UBACyT project 20020250400031BA.

\bibliographystyle{apsrev4-2}
\bibliography{letter_refs}

\begin{thebibliography}{46}%
\makeatletter
\providecommand \@ifxundefined [1]{%
 \@ifx{#1\undefined}
}%
\providecommand \@ifnum [1]{%
 \ifnum #1\expandafter \@firstoftwo
 \else \expandafter \@secondoftwo
 \fi
}%
\providecommand \@ifx [1]{%
 \ifx #1\expandafter \@firstoftwo
 \else \expandafter \@secondoftwo
 \fi
}%
\providecommand \natexlab [1]{#1}%
\providecommand \enquote  [1]{``#1''}%
\providecommand \bibnamefont  [1]{#1}%
\providecommand \bibfnamefont [1]{#1}%
\providecommand \citenamefont [1]{#1}%
\providecommand \href@noop [0]{\@secondoftwo}%
\providecommand \href [0]{\begingroup \@sanitize@url \@href}%
\providecommand \@href[1]{\@@startlink{#1}\@@href}%
\providecommand \@@href[1]{\endgroup#1\@@endlink}%
\providecommand \@sanitize@url [0]{\catcode `\\12\catcode `\$12\catcode `\&12\catcode `\#12\catcode `\^12\catcode `\_12\catcode `\%12\relax}%
\providecommand \@@startlink[1]{}%
\providecommand \@@endlink[0]{}%
\providecommand \url  [0]{\begingroup\@sanitize@url \@url }%
\providecommand \@url [1]{\endgroup\@href {#1}{\urlprefix }}%
\providecommand \urlprefix  [0]{URL }%
\providecommand \Eprint [0]{\href }%
\providecommand \doibase [0]{https://doi.org/}%
\providecommand \selectlanguage [0]{\@gobble}%
\providecommand \bibinfo  [0]{\@secondoftwo}%
\providecommand \bibfield  [0]{\@secondoftwo}%
\providecommand \translation [1]{[#1]}%
\providecommand \BibitemOpen [0]{}%
\providecommand \bibitemStop [0]{}%
\providecommand \bibitemNoStop [0]{.\EOS\space}%
\providecommand \EOS [0]{\spacefactor3000\relax}%
\providecommand \BibitemShut  [1]{\csname bibitem#1\endcsname}%
\let\auto@bib@innerbib\@empty
\bibitem [{\citenamefont {Strominger}\ and\ \citenamefont {Vafa}(1996)}]{Strominger:1996sh}%
  \BibitemOpen
  \bibfield  {author} {\bibinfo {author} {\bibfnamefont {A.}~\bibnamefont {Strominger}}\ and\ \bibinfo {author} {\bibfnamefont {C.}~\bibnamefont {Vafa}},\ }\href@noop {} {\bibfield  {journal} {\bibinfo  {journal} {Phys. Lett. B}\ }\textbf {\bibinfo {volume} {379}},\ \bibinfo {pages} {99} (\bibinfo {year} {1996})},\ \Eprint {https://arxiv.org/abs/hep-th/9601029} {arXiv:hep-th/9601029} \BibitemShut {NoStop}%
\bibitem [{\citenamefont {Banados}\ \emph {et~al.}(1992)\citenamefont {Banados}, \citenamefont {Teitelboim},\ and\ \citenamefont {Zanelli}}]{Banados:1992wn}%
  \BibitemOpen
  \bibfield  {author} {\bibinfo {author} {\bibfnamefont {M.}~\bibnamefont {Banados}}, \bibinfo {author} {\bibfnamefont {C.}~\bibnamefont {Teitelboim}},\ and\ \bibinfo {author} {\bibfnamefont {J.}~\bibnamefont {Zanelli}},\ }\href {https://doi.org/10.1103/PhysRevLett.69.1849} {\bibfield  {journal} {\bibinfo  {journal} {Phys. Rev. Lett.}\ }\textbf {\bibinfo {volume} {69}},\ \bibinfo {pages} {1849} (\bibinfo {year} {1992})},\ \Eprint {https://arxiv.org/abs/hep-th/9204099} {arXiv:hep-th/9204099 [hep-th]} \BibitemShut {NoStop}%
\bibitem [{\citenamefont {Strominger}(1998)}]{Strominger:1997eq}%
  \BibitemOpen
  \bibfield  {author} {\bibinfo {author} {\bibfnamefont {A.}~\bibnamefont {Strominger}},\ }\href@noop {} {\bibfield  {journal} {\bibinfo  {journal} {JHEP}\ }\textbf {\bibinfo {volume} {02}},\ \bibinfo {pages} {009}},\ \Eprint {https://arxiv.org/abs/hep-th/9712251} {arXiv:hep-th/9712251} \BibitemShut {NoStop}%
\bibitem [{\citenamefont {Witten}(1991)}]{Witten:1991yr}%
  \BibitemOpen
  \bibfield  {author} {\bibinfo {author} {\bibfnamefont {E.}~\bibnamefont {Witten}},\ }\href {https://doi.org/10.1103/PhysRevD.44.314} {\bibfield  {journal} {\bibinfo  {journal} {Phys. Rev.}\ }\textbf {\bibinfo {volume} {D44}},\ \bibinfo {pages} {314} (\bibinfo {year} {1991})}\BibitemShut {NoStop}%
\bibitem [{\citenamefont {Martinec}\ \emph {et~al.}(2018)\citenamefont {Martinec}, \citenamefont {Massai},\ and\ \citenamefont {Turton}}]{Martinec:2018nco}%
  \BibitemOpen
  \bibfield  {author} {\bibinfo {author} {\bibfnamefont {E.~J.}\ \bibnamefont {Martinec}}, \bibinfo {author} {\bibfnamefont {S.}~\bibnamefont {Massai}},\ and\ \bibinfo {author} {\bibfnamefont {D.}~\bibnamefont {Turton}},\ }\href {https://doi.org/10.1007/JHEP09(2018)031} {\bibfield  {journal} {\bibinfo  {journal} {JHEP}\ }\textbf {\bibinfo {volume} {09}},\ \bibinfo {pages} {031}},\ \Eprint {https://arxiv.org/abs/1803.08505} {arXiv:1803.08505 [hep-th]} \BibitemShut {NoStop}%
\bibitem [{\citenamefont {Bufalini}\ \emph {et~al.}(2021)\citenamefont {Bufalini}, \citenamefont {Iguri}, \citenamefont {Kovensky},\ and\ \citenamefont {Turton}}]{Bufalini:2021ndn}%
  \BibitemOpen
  \bibfield  {author} {\bibinfo {author} {\bibfnamefont {D.}~\bibnamefont {Bufalini}}, \bibinfo {author} {\bibfnamefont {S.}~\bibnamefont {Iguri}}, \bibinfo {author} {\bibfnamefont {N.}~\bibnamefont {Kovensky}},\ and\ \bibinfo {author} {\bibfnamefont {D.}~\bibnamefont {Turton}},\ }\href {https://doi.org/10.1007/JHEP08(2021)011} {\bibfield  {journal} {\bibinfo  {journal} {JHEP}\ }\textbf {\bibinfo {volume} {08}},\ \bibinfo {pages} {011}},\ \Eprint {https://arxiv.org/abs/2105.02255} {arXiv:2105.02255 [hep-th]} \BibitemShut {NoStop}%
\bibitem [{\citenamefont {Bufalini}\ \emph {et~al.}(2022)\citenamefont {Bufalini}, \citenamefont {Iguri}, \citenamefont {Kovensky},\ and\ \citenamefont {Turton}}]{Bufalini:2022wyp}%
  \BibitemOpen
  \bibfield  {author} {\bibinfo {author} {\bibfnamefont {D.}~\bibnamefont {Bufalini}}, \bibinfo {author} {\bibfnamefont {S.}~\bibnamefont {Iguri}}, \bibinfo {author} {\bibfnamefont {N.}~\bibnamefont {Kovensky}},\ and\ \bibinfo {author} {\bibfnamefont {D.}~\bibnamefont {Turton}},\ }\href {https://doi.org/10.1103/PhysRevLett.129.121603} {\bibfield  {journal} {\bibinfo  {journal} {Phys. Rev. Lett.}\ }\textbf {\bibinfo {volume} {129}},\ \bibinfo {pages} {121603} (\bibinfo {year} {2022})},\ \Eprint {https://arxiv.org/abs/2203.13828} {arXiv:2203.13828 [hep-th]} \BibitemShut {NoStop}%
\bibitem [{\citenamefont {Giveon}\ \emph {et~al.}(2017)\citenamefont {Giveon}, \citenamefont {Itzhaki},\ and\ \citenamefont {Kutasov}}]{Giveon:2017nie}%
  \BibitemOpen
  \bibfield  {author} {\bibinfo {author} {\bibfnamefont {A.}~\bibnamefont {Giveon}}, \bibinfo {author} {\bibfnamefont {N.}~\bibnamefont {Itzhaki}},\ and\ \bibinfo {author} {\bibfnamefont {D.}~\bibnamefont {Kutasov}},\ }\href {https://doi.org/10.1007/JHEP07(2017)122} {\bibfield  {journal} {\bibinfo  {journal} {JHEP}\ }\textbf {\bibinfo {volume} {07}},\ \bibinfo {pages} {122}},\ \Eprint {https://arxiv.org/abs/1701.05576} {arXiv:1701.05576 [hep-th]} \BibitemShut {NoStop}%
\bibitem [{\citenamefont {Chakraborty}\ \emph {et~al.}(2019)\citenamefont {Chakraborty}, \citenamefont {Giveon},\ and\ \citenamefont {Kutasov}}]{Chakraborty:2019mdf}%
  \BibitemOpen
  \bibfield  {author} {\bibinfo {author} {\bibfnamefont {S.}~\bibnamefont {Chakraborty}}, \bibinfo {author} {\bibfnamefont {A.}~\bibnamefont {Giveon}},\ and\ \bibinfo {author} {\bibfnamefont {D.}~\bibnamefont {Kutasov}},\ }\href {https://doi.org/10.1088/1751-8121/ab3710} {\bibfield  {journal} {\bibinfo  {journal} {J. Phys. A}\ }\textbf {\bibinfo {volume} {52}},\ \bibinfo {pages} {384003} (\bibinfo {year} {2019})},\ \Eprint {https://arxiv.org/abs/1905.00051} {arXiv:1905.00051 [hep-th]} \BibitemShut {NoStop}%
\bibitem [{\citenamefont {Dei}\ \emph {et~al.}(2024)\citenamefont {Dei}, \citenamefont {Knighton}, \citenamefont {Naderi},\ and\ \citenamefont {Sethi}}]{Dei:2024sct}%
  \BibitemOpen
  \bibfield  {author} {\bibinfo {author} {\bibfnamefont {A.}~\bibnamefont {Dei}}, \bibinfo {author} {\bibfnamefont {B.}~\bibnamefont {Knighton}}, \bibinfo {author} {\bibfnamefont {K.}~\bibnamefont {Naderi}},\ and\ \bibinfo {author} {\bibfnamefont {S.}~\bibnamefont {Sethi}},\ }\href {https://doi.org/10.1007/JHEP11(2024)145} {\bibfield  {journal} {\bibinfo  {journal} {JHEP}\ }\textbf {\bibinfo {volume} {11}},\ \bibinfo {pages} {145}},\ \Eprint {https://arxiv.org/abs/2408.00823} {arXiv:2408.00823 [hep-th]} \BibitemShut {NoStop}%
\bibitem [{\citenamefont {Kutasov}\ and\ \citenamefont {Seiberg}(1999)}]{Kutasov:1999xu}%
  \BibitemOpen
  \bibfield  {author} {\bibinfo {author} {\bibfnamefont {D.}~\bibnamefont {Kutasov}}\ and\ \bibinfo {author} {\bibfnamefont {N.}~\bibnamefont {Seiberg}},\ }\href {https://doi.org/10.1088/1126-6708/1999/04/008} {\bibfield  {journal} {\bibinfo  {journal} {JHEP}\ }\textbf {\bibinfo {volume} {04}},\ \bibinfo {pages} {008}},\ \Eprint {https://arxiv.org/abs/hep-th/9903219} {arXiv:hep-th/9903219 [hep-th]} \BibitemShut {NoStop}%
\bibitem [{\citenamefont {Maldacena}\ and\ \citenamefont {Ooguri}(2001)}]{Maldacena:2000hw}%
  \BibitemOpen
  \bibfield  {author} {\bibinfo {author} {\bibfnamefont {J.~M.}\ \bibnamefont {Maldacena}}\ and\ \bibinfo {author} {\bibfnamefont {H.}~\bibnamefont {Ooguri}},\ }\href {https://doi.org/10.1063/1.1377273} {\bibfield  {journal} {\bibinfo  {journal} {J. Math. Phys.}\ }\textbf {\bibinfo {volume} {42}},\ \bibinfo {pages} {2929} (\bibinfo {year} {2001})},\ \Eprint {https://arxiv.org/abs/hep-th/0001053} {arXiv:hep-th/0001053 [hep-th]} \BibitemShut {NoStop}%
\bibitem [{\citenamefont {Gaberdiel}\ and\ \citenamefont {Gopakumar}(2018)}]{Gaberdiel:2018rqv}%
  \BibitemOpen
  \bibfield  {author} {\bibinfo {author} {\bibfnamefont {M.~R.}\ \bibnamefont {Gaberdiel}}\ and\ \bibinfo {author} {\bibfnamefont {R.}~\bibnamefont {Gopakumar}},\ }\href {https://doi.org/10.1007/JHEP05(2018)085} {\bibfield  {journal} {\bibinfo  {journal} {JHEP}\ }\textbf {\bibinfo {volume} {05}},\ \bibinfo {pages} {085}},\ \Eprint {https://arxiv.org/abs/1803.04423} {arXiv:1803.04423 [hep-th]} \BibitemShut {NoStop}%
\bibitem [{\citenamefont {Eberhardt}\ \emph {et~al.}(2019)\citenamefont {Eberhardt}, \citenamefont {Gaberdiel},\ and\ \citenamefont {Gopakumar}}]{Eberhardt:2018ouy}%
  \BibitemOpen
  \bibfield  {author} {\bibinfo {author} {\bibfnamefont {L.}~\bibnamefont {Eberhardt}}, \bibinfo {author} {\bibfnamefont {M.~R.}\ \bibnamefont {Gaberdiel}},\ and\ \bibinfo {author} {\bibfnamefont {R.}~\bibnamefont {Gopakumar}},\ }\href {https://doi.org/10.1007/JHEP04(2019)103} {\bibfield  {journal} {\bibinfo  {journal} {JHEP}\ }\textbf {\bibinfo {volume} {04}},\ \bibinfo {pages} {103}},\ \Eprint {https://arxiv.org/abs/1812.01007} {arXiv:1812.01007 [hep-th]} \BibitemShut {NoStop}%
\bibitem [{\citenamefont {Eberhardt}\ \emph {et~al.}(2020)\citenamefont {Eberhardt}, \citenamefont {Gaberdiel},\ and\ \citenamefont {Gopakumar}}]{Eberhardt:2019ywk}%
  \BibitemOpen
  \bibfield  {author} {\bibinfo {author} {\bibfnamefont {L.}~\bibnamefont {Eberhardt}}, \bibinfo {author} {\bibfnamefont {M.~R.}\ \bibnamefont {Gaberdiel}},\ and\ \bibinfo {author} {\bibfnamefont {R.}~\bibnamefont {Gopakumar}},\ }\href {https://doi.org/10.1007/JHEP02(2020)136} {\bibfield  {journal} {\bibinfo  {journal} {JHEP}\ }\textbf {\bibinfo {volume} {2020}}\bibfield  {number} {\bibinfo  {number} { (02)},\ \bibinfo {pages} {136}},\ }\Eprint {https://arxiv.org/abs/1911.00378} {arXiv:1911.00378 [hep-th]} \BibitemShut {NoStop}%
\bibitem [{\citenamefont {Eberhardt}(2022)}]{Eberhardt:2021vsx}%
  \BibitemOpen
  \bibfield  {author} {\bibinfo {author} {\bibfnamefont {L.}~\bibnamefont {Eberhardt}},\ }\href {https://doi.org/10.1088/1751-8121/ac47b2} {\bibfield  {journal} {\bibinfo  {journal} {J. Phys. A}\ }\textbf {\bibinfo {volume} {55}},\ \bibinfo {pages} {064001} (\bibinfo {year} {2022})},\ \Eprint {https://arxiv.org/abs/2110.07535} {arXiv:2110.07535 [hep-th]} \BibitemShut {NoStop}%
\bibitem [{\citenamefont {Balthazar}\ \emph {et~al.}(2022)\citenamefont {Balthazar}, \citenamefont {Giveon}, \citenamefont {Kutasov},\ and\ \citenamefont {Martinec}}]{Balthazar:2021xeh}%
  \BibitemOpen
  \bibfield  {author} {\bibinfo {author} {\bibfnamefont {B.}~\bibnamefont {Balthazar}}, \bibinfo {author} {\bibfnamefont {A.}~\bibnamefont {Giveon}}, \bibinfo {author} {\bibfnamefont {D.}~\bibnamefont {Kutasov}},\ and\ \bibinfo {author} {\bibfnamefont {E.~J.}\ \bibnamefont {Martinec}},\ }\href {https://doi.org/10.1007/JHEP01(2022)008} {\bibfield  {journal} {\bibinfo  {journal} {JHEP}\ }\textbf {\bibinfo {volume} {01}},\ \bibinfo {pages} {008}},\ \Eprint {https://arxiv.org/abs/2109.00065} {arXiv:2109.00065 [hep-th]} \BibitemShut {NoStop}%
\bibitem [{\citenamefont {Maldacena}\ and\ \citenamefont {Ooguri}(2002)}]{Maldacena:2001km}%
  \BibitemOpen
  \bibfield  {author} {\bibinfo {author} {\bibfnamefont {J.~M.}\ \bibnamefont {Maldacena}}\ and\ \bibinfo {author} {\bibfnamefont {H.}~\bibnamefont {Ooguri}},\ }\href {https://doi.org/10.1103/PhysRevD.65.106006} {\bibfield  {journal} {\bibinfo  {journal} {Phys. Rev.}\ }\textbf {\bibinfo {volume} {D65}},\ \bibinfo {pages} {106006} (\bibinfo {year} {2002})},\ \Eprint {https://arxiv.org/abs/hep-th/0111180} {arXiv:hep-th/0111180 [hep-th]} \BibitemShut {NoStop}%
\bibitem [{\citenamefont {Dei}\ and\ \citenamefont {Eberhardt}(2021{\natexlab{a}})}]{Dei:2021xgh}%
  \BibitemOpen
  \bibfield  {author} {\bibinfo {author} {\bibfnamefont {A.}~\bibnamefont {Dei}}\ and\ \bibinfo {author} {\bibfnamefont {L.}~\bibnamefont {Eberhardt}},\ }\href {https://doi.org/10.1007/JHEP08(2021)025} {\bibfield  {journal} {\bibinfo  {journal} {JHEP}\ }\textbf {\bibinfo {volume} {2021}}\bibfield  {number} {\bibinfo  {number} { (08)},\ \bibinfo {pages} {025}},\ }\Eprint {https://arxiv.org/abs/2105.12130} {arXiv:2105.12130 [hep-th]} \BibitemShut {NoStop}%
\bibitem [{\citenamefont {Bufalini}\ \emph {et~al.}(2023)\citenamefont {Bufalini}, \citenamefont {Iguri},\ and\ \citenamefont {Kovensky}}]{Bufalini:2022toj}%
  \BibitemOpen
  \bibfield  {author} {\bibinfo {author} {\bibfnamefont {D.}~\bibnamefont {Bufalini}}, \bibinfo {author} {\bibfnamefont {S.}~\bibnamefont {Iguri}},\ and\ \bibinfo {author} {\bibfnamefont {N.}~\bibnamefont {Kovensky}},\ }\href {https://doi.org/10.1007/JHEP02(2023)246} {\bibfield  {journal} {\bibinfo  {journal} {JHEP}\ }\textbf {\bibinfo {volume} {02}},\ \bibinfo {pages} {246}},\ \Eprint {https://arxiv.org/abs/2212.05877} {arXiv:2212.05877 [hep-th]} \BibitemShut {NoStop}%
\bibitem [{\citenamefont {Dei}\ and\ \citenamefont {Eberhardt}(2021{\natexlab{b}})}]{Dei:2021yom}%
  \BibitemOpen
  \bibfield  {author} {\bibinfo {author} {\bibfnamefont {A.}~\bibnamefont {Dei}}\ and\ \bibinfo {author} {\bibfnamefont {L.}~\bibnamefont {Eberhardt}},\ }\href {https://doi.org/10.1007/JHEP09(2021)209} {\bibfield  {journal} {\bibinfo  {journal} {JHEP}\ }\textbf {\bibinfo {volume} {2021}}\bibfield  {number} {\bibinfo  {number} { (09)},\ \bibinfo {pages} {209}},\ }\Eprint {https://arxiv.org/abs/2107.01481} {arXiv:2107.01481 [hep-th]} \BibitemShut {NoStop}%
\bibitem [{\citenamefont {Hikida}\ and\ \citenamefont {Schomerus}(2024)}]{Hikida:2023jyc}%
  \BibitemOpen
  \bibfield  {author} {\bibinfo {author} {\bibfnamefont {Y.}~\bibnamefont {Hikida}}\ and\ \bibinfo {author} {\bibfnamefont {V.}~\bibnamefont {Schomerus}},\ }\href {https://doi.org/10.1007/JHEP06(2024)071} {\bibfield  {journal} {\bibinfo  {journal} {JHEP}\ }\textbf {\bibinfo {volume} {2024}}\bibfield  {number} {\bibinfo  {number} { (06)},\ \bibinfo {pages} {071}},\ }\Eprint {https://arxiv.org/abs/2312.05317} {arXiv:2312.05317 [hep-th]} \BibitemShut {NoStop}%
\bibitem [{\citenamefont {Knighton}\ \emph {et~al.}(2024)\citenamefont {Knighton}, \citenamefont {Seet},\ and\ \citenamefont {Sriprachyakul}}]{Knighton:2023mhq}%
  \BibitemOpen
  \bibfield  {author} {\bibinfo {author} {\bibfnamefont {B.}~\bibnamefont {Knighton}}, \bibinfo {author} {\bibfnamefont {S.}~\bibnamefont {Seet}},\ and\ \bibinfo {author} {\bibfnamefont {V.}~\bibnamefont {Sriprachyakul}},\ }\href {https://doi.org/10.1007/JHEP05(2024)113} {\bibfield  {journal} {\bibinfo  {journal} {JHEP}\ }\textbf {\bibinfo {volume} {2024}}\bibfield  {number} {\bibinfo  {number} { (05)},\ \bibinfo {pages} {113}},\ }\Eprint {https://arxiv.org/abs/2312.08429} {arXiv:2312.08429 [hep-th]} \BibitemShut {NoStop}%
\bibitem [{\citenamefont {Knighton}\ and\ \citenamefont {Sriprachyakul}(2025)}]{Knighton:2024qxd}%
  \BibitemOpen
  \bibfield  {author} {\bibinfo {author} {\bibfnamefont {B.}~\bibnamefont {Knighton}}\ and\ \bibinfo {author} {\bibfnamefont {V.}~\bibnamefont {Sriprachyakul}},\ }\href {https://doi.org/10.1007/JHEP01(2025)042} {\bibfield  {journal} {\bibinfo  {journal} {JHEP}\ }\textbf {\bibinfo {volume} {2025}}\bibfield  {number} {\bibinfo  {number} { (01)},\ \bibinfo {pages} {042}},\ }\Eprint {https://arxiv.org/abs/2404.07296} {arXiv:2404.07296 [hep-th]} \BibitemShut {NoStop}%
\bibitem [{\citenamefont {Iguri}\ \emph {et~al.}(2026{\natexlab{a}})\citenamefont {Iguri}, \citenamefont {Kovensky},\ and\ \citenamefont {Toro}}]{Iguri:2026lxp}%
  \BibitemOpen
  \bibfield  {author} {\bibinfo {author} {\bibfnamefont {S.}~\bibnamefont {Iguri}}, \bibinfo {author} {\bibfnamefont {N.}~\bibnamefont {Kovensky}},\ and\ \bibinfo {author} {\bibfnamefont {J.~H.}\ \bibnamefont {Toro}},\ }\href@noop {} {\bibinfo {title} {{}}} (\bibinfo {year} {2026}{\natexlab{a}}),\ \Eprint {https://arxiv.org/abs/2609.35495} {arXiv:2609.35495 [hep-th]} \BibitemShut {NoStop}%
\bibitem [{\citenamefont {McStay}\ \emph {et~al.}(2026)\citenamefont {McStay}, \citenamefont {Reid-Edwards},\ and\ \citenamefont {Seet}}]{McStay:2026hlr}%
  \BibitemOpen
  \bibfield  {author} {\bibinfo {author} {\bibfnamefont {N.}~\bibnamefont {McStay}}, \bibinfo {author} {\bibfnamefont {R.}~\bibnamefont {Reid-Edwards}},\ and\ \bibinfo {author} {\bibfnamefont {S.}~\bibnamefont {Seet}},\ }\href@noop {} {\bibinfo {title} {{Spectral flow in AdS$\mathbf{_3}$ revisited}}} (\bibinfo {year} {2026}),\ \Eprint {https://arxiv.org/abs/2611.06295} {arXiv:2611.06295 [hep-th]} \BibitemShut {NoStop}%
\bibitem [{\citenamefont {Iguri}\ \emph {et~al.}(2026{\natexlab{b}})\citenamefont {Iguri}, \citenamefont {Kovensky},\ and\ \citenamefont {Toro}}]{Toro:general}%
  \BibitemOpen
  \bibfield  {author} {\bibinfo {author} {\bibfnamefont {S.}~\bibnamefont {Iguri}}, \bibinfo {author} {\bibfnamefont {N.}~\bibnamefont {Kovensky}},\ and\ \bibinfo {author} {\bibfnamefont {J.~H.}\ \bibnamefont {Toro}},\ }\href@noop {} {\bibinfo {title} {{}}} (\bibinfo {year} {2026}{\natexlab{b}}),\ \bibinfo {note} {to appear}\BibitemShut {NoStop}%
\bibitem [{\citenamefont {Lunin}\ and\ \citenamefont {Mathur}(2001)}]{Lunin:2000yv}%
  \BibitemOpen
  \bibfield  {author} {\bibinfo {author} {\bibfnamefont {O.}~\bibnamefont {Lunin}}\ and\ \bibinfo {author} {\bibfnamefont {S.~D.}\ \bibnamefont {Mathur}},\ }\href {https://doi.org/10.1007/s002200100431} {\bibfield  {journal} {\bibinfo  {journal} {Commun. Math. Phys.}\ }\textbf {\bibinfo {volume} {219}},\ \bibinfo {pages} {399} (\bibinfo {year} {2001})},\ \Eprint {https://arxiv.org/abs/hep-th/0006196} {arXiv:hep-th/0006196} \BibitemShut {NoStop}%
\bibitem [{\citenamefont {Knizhnik}\ and\ \citenamefont {Zamolodchikov}(1984)}]{Knizhnik:1984nr}%
  \BibitemOpen
  \bibfield  {author} {\bibinfo {author} {\bibfnamefont {V.~G.}\ \bibnamefont {Knizhnik}}\ and\ \bibinfo {author} {\bibfnamefont {A.~B.}\ \bibnamefont {Zamolodchikov}},\ }\href {https://doi.org/10.1016/0550-3213(84)90374-2} {\bibfield  {journal} {\bibinfo  {journal} {Nucl. Phys. B}\ }\textbf {\bibinfo {volume} {247}},\ \bibinfo {pages} {83} (\bibinfo {year} {1984})}\BibitemShut {NoStop}%
\bibitem [{\citenamefont {Iguri}\ and\ \citenamefont {Kovensky}(2022)}]{Iguri:2022eat}%
  \BibitemOpen
  \bibfield  {author} {\bibinfo {author} {\bibfnamefont {S.}~\bibnamefont {Iguri}}\ and\ \bibinfo {author} {\bibfnamefont {N.}~\bibnamefont {Kovensky}},\ }\href {https://doi.org/10.21468/SciPostPhys.13.5.115} {\bibfield  {journal} {\bibinfo  {journal} {SciPost Phys.}\ }\textbf {\bibinfo {volume} {13}},\ \bibinfo {pages} {115} (\bibinfo {year} {2022})},\ \Eprint {https://arxiv.org/abs/2208.00978} {arXiv:2208.00978 [hep-th]} \BibitemShut {NoStop}%
\bibitem [{\citenamefont {Fateev}\ \emph {et~al.}()\citenamefont {Fateev}, \citenamefont {Zamolodchikov},\ and\ \citenamefont {Zamolodchikov}}]{Fateev}%
  \BibitemOpen
  \bibfield  {author} {\bibinfo {author} {\bibfnamefont {V.}~\bibnamefont {Fateev}}, \bibinfo {author} {\bibfnamefont {A.}~\bibnamefont {Zamolodchikov}},\ and\ \bibinfo {author} {\bibfnamefont {A.}~\bibnamefont {Zamolodchikov}},\ }\href@noop {} {\bibinfo {title} {{Unpublished notes}}}\BibitemShut {NoStop}%
\bibitem [{\citenamefont {Ribault}\ and\ \citenamefont {Teschner}(2005)}]{Ribault:2005wp}%
  \BibitemOpen
  \bibfield  {author} {\bibinfo {author} {\bibfnamefont {S.}~\bibnamefont {Ribault}}\ and\ \bibinfo {author} {\bibfnamefont {J.}~\bibnamefont {Teschner}},\ }\href {https://doi.org/10.1088/1126-6708/2005/06/014} {\bibfield  {journal} {\bibinfo  {journal} {JHEP}\ }\textbf {\bibinfo {volume} {06}},\ \bibinfo {pages} {014}},\ \Eprint {https://arxiv.org/abs/hep-th/0502048} {arXiv:hep-th/0502048} \BibitemShut {NoStop}%
\bibitem [{\citenamefont {Ribault}(2005)}]{Ribault:2005ms}%
  \BibitemOpen
  \bibfield  {author} {\bibinfo {author} {\bibfnamefont {S.}~\bibnamefont {Ribault}},\ }\href {https://doi.org/10.1088/1126-6708/2005/09/045} {\bibfield  {journal} {\bibinfo  {journal} {JHEP}\ }\textbf {\bibinfo {volume} {2005}}\bibfield  {number} {\bibinfo  {number} { (09)},\ \bibinfo {pages} {045}},\ }\Eprint {https://arxiv.org/abs/hep-th/0507114} {arXiv:hep-th/0507114} \BibitemShut {NoStop}%
\bibitem [{\citenamefont {Dei}\ and\ \citenamefont {Eberhardt}(2022)}]{Dei:2022pkr}%
  \BibitemOpen
  \bibfield  {author} {\bibinfo {author} {\bibfnamefont {A.}~\bibnamefont {Dei}}\ and\ \bibinfo {author} {\bibfnamefont {L.}~\bibnamefont {Eberhardt}},\ }\href {https://doi.org/10.21468/SciPostPhys.13.3.053} {\bibfield  {journal} {\bibinfo  {journal} {SciPost Phys.}\ }\textbf {\bibinfo {volume} {13}},\ \bibinfo {pages} {053} (\bibinfo {year} {2022})},\ \Eprint {https://arxiv.org/abs/2203.13264} {arXiv:2203.13264 [hep-th]} \BibitemShut {NoStop}%
\bibitem [{\citenamefont {Teschner}(1999)}]{Teschner:1997ft}%
  \BibitemOpen
  \bibfield  {author} {\bibinfo {author} {\bibfnamefont {J.}~\bibnamefont {Teschner}},\ }\href {https://doi.org/10.1016/S0550-3213(99)00072-3} {\bibfield  {journal} {\bibinfo  {journal} {Nucl. Phys.}\ }\textbf {\bibinfo {volume} {B546}},\ \bibinfo {pages} {390} (\bibinfo {year} {1999})},\ \Eprint {https://arxiv.org/abs/hep-th/9712256} {arXiv:hep-th/9712256 [hep-th]} \BibitemShut {NoStop}%
\bibitem [{\citenamefont {Teschner}(2000)}]{Teschner:1999ug}%
  \BibitemOpen
  \bibfield  {author} {\bibinfo {author} {\bibfnamefont {J.}~\bibnamefont {Teschner}},\ }\href {https://doi.org/10.1016/S0550-3213(99)00785-3} {\bibfield  {journal} {\bibinfo  {journal} {Nucl. Phys.}\ }\textbf {\bibinfo {volume} {B571}},\ \bibinfo {pages} {555} (\bibinfo {year} {2000})},\ \Eprint {https://arxiv.org/abs/hep-th/9906215} {arXiv:hep-th/9906215 [hep-th]} \BibitemShut {NoStop}%
\bibitem [{\citenamefont {Iguri}\ \emph {et~al.}(2024)\citenamefont {Iguri}, \citenamefont {Kovensky},\ and\ \citenamefont {Toro}}]{Iguri:2024yhb}%
  \BibitemOpen
  \bibfield  {author} {\bibinfo {author} {\bibfnamefont {S.}~\bibnamefont {Iguri}}, \bibinfo {author} {\bibfnamefont {N.}~\bibnamefont {Kovensky}},\ and\ \bibinfo {author} {\bibfnamefont {J.~H.}\ \bibnamefont {Toro}},\ }\href {https://doi.org/10.1007/JHEP08(2024)186} {\bibfield  {journal} {\bibinfo  {journal} {JHEP}\ }\textbf {\bibinfo {volume} {2024}}\bibfield  {number} {\bibinfo  {number} { (08)},\ \bibinfo {pages} {186}},\ }\Eprint {https://arxiv.org/abs/2406.04247} {arXiv:2406.04247 [hep-th]} \BibitemShut {NoStop}%
\bibitem [{\citenamefont {Sriprachyakul}(2025)}]{Sriprachyakul:2025ubx}%
  \BibitemOpen
  \bibfield  {author} {\bibinfo {author} {\bibfnamefont {V.}~\bibnamefont {Sriprachyakul}},\ }\href {https://doi.org/10.1007/JHEP08(2025)159} {\bibfield  {journal} {\bibinfo  {journal} {JHEP}\ }\textbf {\bibinfo {volume} {08}},\ \bibinfo {pages} {159}},\ \Eprint {https://arxiv.org/abs/2504.18244} {arXiv:2504.18244 [hep-th]} \BibitemShut {NoStop}%
\bibitem [{\citenamefont {Seiberg}\ and\ \citenamefont {Witten}(1999)}]{Seiberg:1999xz}%
  \BibitemOpen
  \bibfield  {author} {\bibinfo {author} {\bibfnamefont {N.}~\bibnamefont {Seiberg}}\ and\ \bibinfo {author} {\bibfnamefont {E.}~\bibnamefont {Witten}},\ }\href {https://doi.org/10.1088/1126-6708/1999/04/017} {\bibfield  {journal} {\bibinfo  {journal} {JHEP}\ }\textbf {\bibinfo {volume} {04}},\ \bibinfo {pages} {017}},\ \Eprint {https://arxiv.org/abs/hep-th/9903224} {arXiv:hep-th/9903224 [hep-th]} \BibitemShut {NoStop}%
\bibitem [{\citenamefont {Eberhardt}\ and\ \citenamefont {Gaberdiel}(2019)}]{Eberhardt:2019qcl}%
  \BibitemOpen
  \bibfield  {author} {\bibinfo {author} {\bibfnamefont {L.}~\bibnamefont {Eberhardt}}\ and\ \bibinfo {author} {\bibfnamefont {M.~R.}\ \bibnamefont {Gaberdiel}},\ }\href {https://doi.org/10.1016/j.nuclphysb.2019.114774} {\bibfield  {journal} {\bibinfo  {journal} {Nucl. Phys.}\ }\textbf {\bibinfo {volume} {B948}},\ \bibinfo {pages} {114774} (\bibinfo {year} {2019})},\ \Eprint {https://arxiv.org/abs/1903.00421} {arXiv:1903.00421 [hep-th]} \BibitemShut {NoStop}%
\bibitem [{\citenamefont {Zamolodchikov}\ and\ \citenamefont {Zamolodchikov}(1996)}]{Zamolodchikov:1995aa}%
  \BibitemOpen
  \bibfield  {author} {\bibinfo {author} {\bibfnamefont {A.~B.}\ \bibnamefont {Zamolodchikov}}\ and\ \bibinfo {author} {\bibfnamefont {A.~B.}\ \bibnamefont {Zamolodchikov}},\ }\href {https://doi.org/10.1016/0550-3213(96)00351-3} {\bibfield  {journal} {\bibinfo  {journal} {Nucl. Phys. B}\ }\textbf {\bibinfo {volume} {477}},\ \bibinfo {pages} {577} (\bibinfo {year} {1996})},\ \Eprint {https://arxiv.org/abs/hep-th/9506136} {arXiv:hep-th/9506136} \BibitemShut {NoStop}%
\bibitem [{\citenamefont {Iguri}\ \emph {et~al.}(2023)\citenamefont {Iguri}, \citenamefont {Kovensky},\ and\ \citenamefont {Toro}}]{Iguri:2022pbp}%
  \BibitemOpen
  \bibfield  {author} {\bibinfo {author} {\bibfnamefont {S.}~\bibnamefont {Iguri}}, \bibinfo {author} {\bibfnamefont {N.}~\bibnamefont {Kovensky}},\ and\ \bibinfo {author} {\bibfnamefont {J.~H.}\ \bibnamefont {Toro}},\ }\href {https://doi.org/10.1007/JHEP01(2023)161} {\bibfield  {journal} {\bibinfo  {journal} {JHEP}\ }\textbf {\bibinfo {volume} {2023}}\bibfield  {number} {\bibinfo  {number} { (01)},\ \bibinfo {pages} {161}},\ }\Eprint {https://arxiv.org/abs/2211.02521} {arXiv:2211.02521 [hep-th]} \BibitemShut {NoStop}%
\bibitem [{\citenamefont {Barone}\ \emph {et~al.}(2026)\citenamefont {Barone}, \citenamefont {Iguri}, \citenamefont {Kovensky},\ and\ \citenamefont {Toro}}]{Barone:2025vww}%
  \BibitemOpen
  \bibfield  {author} {\bibinfo {author} {\bibfnamefont {E.}~\bibnamefont {Barone}}, \bibinfo {author} {\bibfnamefont {S.}~\bibnamefont {Iguri}}, \bibinfo {author} {\bibfnamefont {N.}~\bibnamefont {Kovensky}},\ and\ \bibinfo {author} {\bibfnamefont {J.~H.}\ \bibnamefont {Toro}},\ }\href {https://doi.org/10.1007/JHEP03(2026)119} {\bibfield  {journal} {\bibinfo  {journal} {JHEP}\ }\textbf {\bibinfo {volume} {2026}}\bibfield  {number} {\bibinfo  {number} { (03)},\ \bibinfo {pages} {119}},\ }\Eprint {https://arxiv.org/abs/2510.15732} {arXiv:2510.15732 [hep-th]} \BibitemShut {NoStop}%
\bibitem [{\citenamefont {Knighton}\ \emph {et~al.}(2026)\citenamefont {Knighton}, \citenamefont {McStay},\ and\ \citenamefont {Sriprachyakul}}]{Knighton:2026wva}%
  \BibitemOpen
  \bibfield  {author} {\bibinfo {author} {\bibfnamefont {B.}~\bibnamefont {Knighton}}, \bibinfo {author} {\bibfnamefont {N.}~\bibnamefont {McStay}},\ and\ \bibinfo {author} {\bibfnamefont {V.}~\bibnamefont {Sriprachyakul}},\ }\href@noop {} {\bibinfo {title} {{}}} (\bibinfo {year} {2026}),\ \Eprint {https://arxiv.org/abs/2603.18142} {arXiv:2603.18142 [hep-th]} \BibitemShut {NoStop}%
\bibitem [{\citenamefont {Hikida}\ and\ \citenamefont {Schomerus}(2009)}]{Hikida:2008pe}%
  \BibitemOpen
  \bibfield  {author} {\bibinfo {author} {\bibfnamefont {Y.}~\bibnamefont {Hikida}}\ and\ \bibinfo {author} {\bibfnamefont {V.}~\bibnamefont {Schomerus}},\ }\href {https://doi.org/10.1088/1126-6708/2009/03/095} {\bibfield  {journal} {\bibinfo  {journal} {JHEP}\ }\textbf {\bibinfo {volume} {03}},\ \bibinfo {pages} {095}},\ \Eprint {https://arxiv.org/abs/0805.3931} {arXiv:0805.3931 [hep-th]} \BibitemShut {NoStop}%
\bibitem [{\citenamefont {Chen}\ \emph {et~al.}(2023)\citenamefont {Chen}, \citenamefont {Maldacena},\ and\ \citenamefont {Witten}}]{Chen:2021dsw}%
  \BibitemOpen
  \bibfield  {author} {\bibinfo {author} {\bibfnamefont {Y.}~\bibnamefont {Chen}}, \bibinfo {author} {\bibfnamefont {J.}~\bibnamefont {Maldacena}},\ and\ \bibinfo {author} {\bibfnamefont {E.}~\bibnamefont {Witten}},\ }\href {https://doi.org/10.1007/JHEP01(2023)103} {\bibfield  {journal} {\bibinfo  {journal} {JHEP}\ }\textbf {\bibinfo {volume} {01}},\ \bibinfo {pages} {103}},\ \Eprint {https://arxiv.org/abs/2109.08563} {arXiv:2109.08563 [hep-th]} \BibitemShut {NoStop}%
\end{thebibliography}%

\section*{End Matter}
\vspace{1.5mm}\noindent Here we prove Eqs.~\eqref{L: motion} and \eqref{L: tau} and check the edge case for $n=4$. Throughout, $i$ labels the operator whose insertion point is varied, $\ell$ any operator, and $f=p-x_iq$ for any pair $(p,q)$. We fix a basis $e_a=(p_a,q_a)$, $a=1,2$, of the pencil that depends smoothly on the insertion points, and take derivatives at fixed $s$, i.e.~$\partial p_s=s\,\partial p_1+\partial p_2$. The polynomials $\varphi_i$ and $\psi_i$ depend on the basis, while $W$ and $\det K$ do not. By Eq.~\eqref{L: area form}, $\partial_s\Gamma_s=R/q_s^2$, so that
\begin{align}
\chi(s;z)&=\frac{1}{\partial_s\Gamma_s(z)}=\frac{q_s(z)^2}{R(z)}\,,\nn\\
K(s;z)&=\frac{p_s'(z)\,q_s(z)-p_s(z)\,q_s'(z)}{R(z)}\,.
\label{E: chi K}
\end{align}

\vspace{1.5mm}\noindent{\bf \textit{Motion of the points}}.---Let $v$ be $x_i$ or $z_i$. The motion of all members is encoded in the velocity
\beq
K_v(s;z)=\frac{\partial_v\Gamma_s}{\partial_s\Gamma_s}=\frac{q_s\,\partial_vp_s-p_s\,\partial_vq_s}{R}\,.
\label{E: velocity}
\eeq
Since $\partial_vp_s$ and $\partial_vq_s$ are polynomials of degree at most $N+1$, linear in $s$, $K_v$ is quadratic in $s$, and, as a function of $z$, can only have poles at the zeros $z_\ell$ of $R$. Its numerator has degree at most $2N+2=\deg R$, so $K_v$ is bounded at infinity. It is therefore fixed by its principal part up to a $z$-independent term. On the other hand, $\xi_\ell$ is defined by $a_\ell(\xi_\ell)=y_\ell$. Differentiating at fixed $y_\ell$ gives
\beq
\partial_v\xi_\ell=-\partial_va_\ell/a_\ell'\big|_{s=\xi_\ell}\,.
\label{E: member motion}
\eeq
To relate the two, we use $\Gamma_s=x_\ell+(z-z_\ell)^{\w_\ell}A_\ell(s;z)$, so that $\partial_s\Gamma_s=(z-z_\ell)^{\w_\ell}\partial_sA_\ell$, with $\partial_sA_\ell\neq0$ at $z_\ell$.

\textit{(a) $v=x_i$.} For $\ell\neq i$, $x_i$ only enters through $A_\ell$. Hence $K_{x_i}=\partial_{x_i}A_\ell/\partial_sA_\ell$ which is regular at $z_\ell$, with value $\partial_{x_i}a_\ell/a_\ell'$. At $z_i$, $x_i$ also enters explicitly, $\partial_{x_i}\Gamma_s=1+(z-z_i)^{\w_i}\partial_{x_i}A_i$, and dividing by $\partial_s\Gamma_s$ gives $K_{x_i}=\chi+\partial_{x_i}A_i/\partial_sA_i$. Thus $K_{x_i}$ has the same pole as $\chi$ at $z_i$, and $K_{x_i}-\chi$ takes the value $\partial_{x_i}a_i/a_i'$ there. Let
\beq
\chi_i^{\rm sing}(s;z)=\Res_{z'=z_i}\frac{\chi(s;z')}{z-z'}
\eeq
be the principal part of $\chi$ at $z_i$. Then $K_{x_i}-\chi_i^{\rm sing}$ is regular at all the $z_\ell$ and bounded at infinity, hence independent of $z$, and
\beq
K_{x_i}(s;z)=\chi_i^{\rm sing}(s;z)-\varphi_i(s)\,,
\label{E: Kx}
\eeq
with $\varphi_i$ quadratic in $s$. We now evaluate Eq.~\eqref{E: member motion}. For $\ell\neq i$, $\partial_{x_i}\xi_\ell=-K_{x_i}(\xi_\ell;z_\ell)$, and $\chi_i^{\rm sing}(z_\ell)=-\Res_{z=z_i}\chi(z)/(z-z_\ell)$. For $\ell=i$, $\partial_{x_i}\xi_i=-(K_{x_i}-\chi)(\xi_i;z_i)=(\chi-\chi_i^{\rm sing})(\xi_i;z_i)+\varphi_i(\xi_i)$, and the regular part $\chi-\chi_i^{\rm sing}$ takes the value $\Res_{z=z_i}\chi(z)/(z-z_i)$ at $z_i$. Both cases thus give Eq.~\eqref{L: motion x}.

\textit{(b) $v=z_i$.} The difference is that $z_i$ also enters through $(z-z_i)^{\w_i}$, on which $\partial_{z_i}$ and $\partial_z$ act with opposite signs. Near $z_i$,
\beq
\partial_{z_i}\Gamma_s+\partial_z\Gamma_s=(z-z_i)^{\w_i}\big(\partial_{z_i}A_i+\partial_zA_i\big)\,,
\eeq
so $K_{z_i}+K$ is regular at $z_i$, with value $\partial_{z_i}a_i/a_i'$, since $a_i(s)=A_i(s;z_i)$ depends on $z_i$ through both arguments. As in (a), $K_{z_i}$ is regular at $z_\ell\neq z_i$, with value $\partial_{z_i}a_\ell/a_\ell'$. Let $K_i^{\rm sing}$ be the principal part of $K$ at $z_i$, defined as for $\chi$. Then $K_{z_i}+K_i^{\rm sing}$ is regular everywhere and bounded, so $K_{z_i}=-K_i^{\rm sing}+\psi_i(s)$, with $\psi_i$ quadratic in $s$. Evaluating at $z_\ell$ as in (a) then gives Eq.~\eqref{L: motion z}.

\vspace{1.5mm}\noindent{\bf \textit{Determinant identities}}.---An unknown of the system~\eqref{L: linear system} at degree $N$ is a pair $\varepsilon=(p_\varepsilon,q_\varepsilon)$ of polynomials of degree at most $N$, and we write $f_\varepsilon=p_\varepsilon-x_iq_\varepsilon$. Each condition is a linear functional of $\varepsilon$, $L_{\ell m}(\varepsilon)=[p_\varepsilon-x_\ell q_\varepsilon]_m(z_\ell)$. The matrix $M$ of the system has these functionals as rows, acting on the coefficients of $p_\varepsilon$ and $q_\varepsilon$, so that $\tau_{\Bw}=\det M$, which we assume non-zero. The columns of $M^{-1}$ are the dual pairs $\varepsilon_{\ell m}$, defined by $L_{\ell'm'}(\varepsilon_{\ell m})=\delta_{\ell\ell'}\delta_{mm'}$. Jacobi's formula then reads
\beq
\partial\log\tau_{\Bw}={\rm tr}\big(M^{-1}\partial M\big)=\sum_{\ell,m}(\partial L_{\ell m})(\varepsilon_{\ell m})\,,
\label{E: Jacobi}
\eeq
where $\partial L_{\ell m}$ is the derivative of the functional at fixed coefficients. Only the $\w_i$ functionals at $z_i$ depend on $x_i$ and $z_i$, and we write $\varepsilon_m\equiv\varepsilon_{im}$ for their dual pairs, $m=0,\dots,\w_i-1$. By definition, $\varepsilon_m$ satisfies all the conditions at the other points, and $f_{\varepsilon_m}=(z-z_i)^m+\cO[(z-z_i)^{\w_i}]$.

The important observation is that multiplying these pairs by $z-z_i$ maps each one to the next, up to a member of the pencil. Indeed, this multiplication preserves the conditions at the other points, and $f_{(z-z_i)\varepsilon_m}$ agrees with $f_{\varepsilon_{m+1}}$ up to $\cO[(z-z_i)^{\w_i}]$. Hence the pairs of degree $N+1$
\beq
(z-z_i)\,\varepsilon_m-\varepsilon_{m+1}\,,\qquad \varepsilon_{\w_i}\equiv0\,,
\label{E: shift}
\eeq
satisfy Eq.~\eqref{L: linear system}, i.e.~they are linear combinations of $e_1$ and $e_2$. We write $f_a=p_a-x_iq_a$.

\textit{Eq.~\eqref{L: tau x}.} Since $\partial_{x_i}L_{im}(\varepsilon)=-[q_\varepsilon]_m(z_i)$, Eq.~\eqref{E: Jacobi} gives
\begin{align}
\partial_{x_i}\log\tau_{\Bw}&=-\sum_{m<\w_i}[q_{\varepsilon_m}]_m(z_i)\nn\\
&=-\Res_{z=z_i}\sum_{m<\w_i}\frac{q_{\varepsilon_m}(z)}{(z-z_i)^{m+1}}\,.
\label{E: dx tau}
\end{align}
To evaluate the sum, we combine the members~\eqref{E: shift} with weights $(z_0-z_i)^{-m-1}$, where $z_0$ is an auxiliary point. Writing $z-z_i=(z-z_0)+(z_0-z_i)$, the sum reads
\begin{align}
V(z)&=\sum_{m<\w_i}\frac{(z-z_i)\,\varepsilon_m(z)-\varepsilon_{m+1}(z)}{(z_0-z_i)^{m+1}}\nn\\
&=(z-z_0)\sum_{m<\w_i}\frac{\varepsilon_m(z)}{(z_0-z_i)^{m+1}}+\varepsilon_0(z)\,.
\end{align}
Thus $V(z_0)=\varepsilon_0(z_0)$, while the $q$-component of $V'(z_0)-\varepsilon_0'(z_0)$ is the sum in Eq.~\eqref{E: dx tau}, at $z=z_0$. On the other hand, $V=c_1e_1+c_2e_2$, with coefficients depending on $z_0$. The condition $V(z_0)=\varepsilon_0(z_0)$ fixes them. In components $(f,q)$ the determinant is $f_1q_2-f_2q_1=R$, and Cramer's rule gives
\begin{align}
c_1&=\frac{f_{\varepsilon_0}q_2-f_2\,q_{\varepsilon_0}}{R}\,,\qquad
c_2=\frac{f_1\,q_{\varepsilon_0}-f_{\varepsilon_0}q_1}{R}\,,
\end{align}
evaluated at $z_0$. The $q$-component of $V'(z_0)=c_1e_1'(z_0)+c_2e_2'(z_0)$ then gives, after renaming $z_0\to z$,
\begin{align}
\sum_{m<\w_i}\frac{q_{\varepsilon_m}(z)}{(z-z_i)^{m+1}}={}&\frac{f_{\varepsilon_0}\,(q_1'q_2-q_1q_2')+q_{\varepsilon_0}\,(f_1q_2'-f_2q_1')}{R}\nn\\
&-q_{\varepsilon_0}'\,.
\label{E: qU}
\end{align}
The first term is $-f_{\varepsilon_0}W$, with $f_{\varepsilon_0}=1+\cO[(z-z_i)^{\w_i}]$, and since $W$ has a pole of order at most $\w_i$ at $z_i$, its residue is $-\Res_{z=z_i}W$. The second term is regular at $z_i$, since $f_a=\cO[(z-z_i)^{\w_i}]$ and $R$ vanishes to order $\w_i$, and the last one is a polynomial. Hence $\partial_{x_i}\log\tau_{\Bw}=\Res_{z=z_i}W$. The prefactor in Eq.~\eqref{L: P} does not depend on $\Bx$, which proves Eq.~\eqref{L: tau x}.

\textit{Eq.~\eqref{L: tau z}.} At fixed coefficients, $\partial_{z_i}L_{im}(\varepsilon)=(m+1)[f_\varepsilon]_{m+1}(z_i)$. On $\varepsilon_m$ this vanishes unless $m=\w_i-1$, since $f_{\varepsilon_m}$ has no terms of order $m+1<\w_i$. Hence Eq.~\eqref{E: Jacobi} gives
\beq
\partial_{z_i}\log\tau_{\Bw}=\w_i\,\rho_i\,,\qquad \rho_i\equiv[f_{\varepsilon_{\w_i-1}}]_{\w_i}(z_i)\,.
\eeq
To relate $\rho_i$ to the pencil, we choose $e_1=(z-z_i)\,\varepsilon_{\w_i-1}$, i.e.~the member~\eqref{E: shift} with $m=\w_i-1$. Its $q_1$ vanishes at $z_i$ and, by the definition of $\rho_i$,
\beq
f_1=(z-z_i)^{\w_i}\big(1+\rho_i\,(z-z_i)+\dots\big)\,.
\eeq
For $e_2$ we take the member whose $f_2$ vanishes to order $\w_i+1$, obtained from any other member by subtracting a multiple of $e_1$. Then $f_2q_1$ vanishes to order $\w_i+2$, and
\beq
\frac{R}{f_1}=q_2-\frac{f_2\,q_1}{f_1}=q_2+\cO[(z-z_i)^2]\,.
\eeq
On the other hand, expanding Eq.~\eqref{L: area form} around $z_i$, $R=C\pi_i\,(z-z_i)^{\w_i}\big(1+\Sigma_i(z-z_i)+\dots\big)$ with $\Sigma_i\equiv\sum_{\ell\neq i}\w_\ell/z_{i\ell}$, so that
\beq
q_2=C\pi_i\big(1+(\Sigma_i-\rho_i)(z-z_i)\big)+\cO[(z-z_i)^2]\,.
\label{E: R expansion}
\eeq
The numerator of $K$ in Eq.~\eqref{E: chi K} is quadratic in $s$, with coefficients $RK^+=p_1'q_1-p_1q_1'$, $RK^-=p_2'q_2-p_2q_2'$ and $-2RK^3=p_1'q_2+p_2'q_1-p_1q_2'-p_2q_1'$. A direct computation, using $R=p_1q_2-p_2q_1$, gives
\beq
\det K=\frac{f_1'q_2'-f_2'q_1'}{R}-\frac14\Big(\frac{R'}{R}\Big)^2\,,
\label{E: detK}
\eeq
where we used $p_1'q_2'-p_2'q_1'=f_1'q_2'-f_2'q_1'$. Since $f_2'$ vanishes to order $\w_i$, only $f_1'q_2'/R=\w_i\,q_2'(z_i)/\big(C\pi_i(z-z_i)\big)+\dots$ contributes to the residue of the first term, which by Eq.~\eqref{E: R expansion} is $\w_i(\Sigma_i-\rho_i)$. In the second, $R'/R=\w_i/(z-z_i)+\Sigma_i+\dots$, so its residue is $-\frac12\w_i\Sigma_i$. Altogether, $\Res_{z=z_i}\det K=\frac12\w_i\Sigma_i-\w_i\rho_i$. Since the prefactor in Eq.~\eqref{L: P} contributes $-\frac12\w_i\Sigma_i$ to $\partial_{z_i}\log\cP_{\Bw}$, this is Eq.~\eqref{L: tau z}.

\vspace{1.5mm}\noindent {\bf \textit{Edge cases for $n=4$}}.---As a check, we compare our result with the edge-case formula of~\cite{Dei:2021yom}, for which $d=2$. We impose Eq.~\eqref{L: local yr} for the operators $V_1$ and $V_3$ without loss of generality, and denote by $\cP_{\{1,3\}}$ the corresponding function \eqref{L: P}. Restoring the antiholomorphic dependence, $\Phi$ is the unflowed correlator $\tilde G$ with spins $(\tilde\jmath_1,j_2,\tilde\jmath_3,j_4)$, worldsheet insertions $(0,1,\infty,z)$ and boundary insertions $(\xi_1,\xi_2,\xi_3,\xi_4)$, evaluated at $\xi_1=\xi_3=\xi_*$, with $\tilde G$ defined in the frame $(0,1,\infty,x)$. The M\"obius map sending $(\xi_1,\xi_2,\xi_3)$ to $(0,1,\infty)$ brings
\beq
\xi_4 \to x'=\frac{\xi_{41}\xi_{23}}{\xi_{43}\xi_{21}} \, ,
\eeq
so that $\xi_1\to\xi_3$ corresponds to $x'\to1$. The covariance factor contains $|\xi_{13}|^{2(J-k)}$, with $J=\sum_ij_i$, hence the coincidence limit selects the term of $\tilde G$ that behaves as $|1-x'|^{2(k-J)}$ \cite{Maldacena:2001km}. Denoting by $g(z)$ its coefficient, this gives
\begin{equation}
\Phi=g(z)\,|\xi_{24}^{k-J}\xi_{2*}^{J-2j_2-k}\xi_{4*}^{J-2j_4-k}|^2 \,.
\end{equation}
On the other hand, the flip identity of~\cite{Dei:2022pkr} expresses the correlator $G(x)$ with the original spins at $\Bx=(0,1,\infty,x)$ as $G(x)=\cN(j_1)\cN(j_3)|x|^{-4j_4}|z|^{2j_4}\tilde G(z/x)$. The limit $x'\to1$ translates into $x\to z$, i.e.
\beq
g(z)=\frac{|z|^{2j_4+2(k-J)}}{\cN(j_1)\cN(j_3)}\,\lim_{x\to z}|x-z|^{2(J-k)}\,G(x) \, .
\eeq
The limit gives the function $f(z)$ of~\cite{Dei:2021yom}. The factors $\cN(j_1)\cN(j_3)$ cancel against those in our result, see Eq.~\eqref{L: odd solution}. As in the odd case, one defines $X_{(ik)}=\sqrt{z_{ik}}\,\cP_{\{i,k\}\triangle \{1,3\}}$, where $\triangle$ denotes the symmetric difference, and with $\cP_I$ given by Eq.~\eqref{L: P} for the charges $\Bw+\sum_{a\in I}e_a$ and the conditions $a_a=y_a$ for $a\in I$. This gives $X_{(ik)}=\cP_{\{1,3\}}\,\xi_{ik}/\sqrt{\xi_i'\xi_k'}$ up to a phase. Including the corresponding exponents, this turns the remaining prefactors into the $X_I$ of~\cite{Dei:2021yom}, while the powers of $z$ cancel. By~\cite{Ribault:2005wp}, $f(z)$ is a Liouville four-point function~\cite{Dei:2022pkr}.
\end{document}